\documentclass{article}

\usepackage{arxiv}

\usepackage{wrapfig}
\usepackage[frozencache,cachedir=.]{minted}
\usepackage{float}
\usepackage{graphicx}
\usepackage{natbib}
\usepackage{amsmath}

\usepackage{caption}
\usepackage{newfloat}

\usepackage{cuted,tcolorbox} 

\usepackage[utf8]{inputenc} 
\usepackage[T1]{fontenc}    
\usepackage{url}            
\usepackage{booktabs}       
\usepackage{amsfonts}       
\usepackage{nicefrac}       
\usepackage{microtype}      
\usepackage{lipsum}

\SetupFloatingEnvironment{listing}{name=Example Code}

\definecolor{DesatCyan}{HTML}{6CBAB8}
\definecolor{DesatMagenta}{HTML}{B86CBA}
\definecolor{DesatMustard}{HTML}{BAB86C}

\usepackage{scalerel}
\usepackage{tikz}
\usetikzlibrary{svg.path}

\definecolor{orcidlogocol}{HTML}{A6CE39}
\tikzset{
  orcidlogo/.pic={
    \fill[orcidlogocol] svg{M256,128c0,70.7-57.3,128-128,128C57.3,256,0,198.7,0,128C0,57.3,57.3,0,128,0C198.7,0,256,57.3,256,128z};
    \fill[white] svg{M86.3,186.2H70.9V79.1h15.4v48.4V186.2z}
                 svg{M108.9,79.1h41.6c39.6,0,57,28.3,57,53.6c0,27.5-21.5,53.6-56.8,53.6h-41.8V79.1z M124.3,172.4h24.5c34.9,0,42.9-26.5,42.9-39.7c0-21.5-13.7-39.7-43.7-39.7h-23.7V172.4z}
                 svg{M88.7,56.8c0,5.5-4.5,10.1-10.1,10.1c-5.6,0-10.1-4.6-10.1-10.1c0-5.6,4.5-10.1,10.1-10.1C84.2,46.7,88.7,51.3,88.7,56.8z};
  }
}

\newcommand\orcidicon[1]{\href{https://orcid.org/#1}{\mbox{\scalerel*{
\begin{tikzpicture}[yscale=-1,transform shape]
\pic{orcidlogo};
\end{tikzpicture}
}{|}}}}

\usepackage[hidelinks]{hyperref} 

\graphicspath{{img/}}

\title{70 years of machine learning in geoscience in review}

\author{
  Jesper S\"oren Dramsch\textsuperscript{\orcidicon{0000-0001-8273-905X}} \\
  \texttt{\href{mailto://jesper@dramsch.net}{jesper@dramsch.net}}
}

\begin{document}
\maketitle


\vspace*{18pt}

\begin{abstract}
This review gives an overview of the development of machine learning in geoscience. A thorough analysis of the co-developments of machine learning applications throughout the last 70 years relates the recent enthusiasm for machine learning to developments in geoscience. I explore the shift of kriging towards a mainstream machine learning method and the historic application of neural networks in geoscience, following the general trend of machine learning enthusiasm through the decades. Furthermore, this chapter explores the shift from mathematical fundamentals and knowledge in software development towards skills in model validation, applied statistics, and integrated subject matter expertise. The review is interspersed with code examples to complement the theoretical foundations and illustrate model validation and machine learning explainability for science. The scope of this review includes various shallow machine learning methods, e.g. Decision Trees, Random Forests, Support-Vector Machines, and Gaussian Processes, as well as, deep neural networks, including feed-forward neural networks, convolutional neural networks, recurrent neural networks and generative adversarial networks. Regarding geoscience, the review has a bias towards geophysics but aims to strike a balance with geochemistry, geostatistics, and geology, however excludes remote sensing, as this would exceed the scope. In general, I aim to provide context for the recent enthusiasm surrounding deep learning with respect to research, hardware, and software developments that enable successful application of shallow and deep machine learning in all disciplines of Earth science. 
\end{abstract}


In recent years machine learning has become an increasingly important interdisciplinary tool that has advanced several fields of science, such as biology \citep{Ching2018-hg}, chemistry \citep{Schutt2017-sh}, medicine \citep{Shen2017-nt} and pharmacology \citep{Kadurin2017-oq}. Specifically, the method of deep neural networks has found wide application. While geoscience was slower in the adoption, bibliometrics show the adoption of deep learning in all aspects of geoscience. Most subdisciplines of geoscience have been treated to a review of machine learning. Remote sensing has been an early adopter \citep{Lary2016-aa}, with geomorphology \citep{valentine2016introduction}, solid Earth geoscience \citep{bergen2019machine}, hydrogeophysics \citep{shen2018transdisciplinary}, seismology \citep{kong2019machine}, seismic interpretation \citep{wang2018successful} and geochemistry \citep{zuo2019deep} following suite. Climate change, in particular, has received a thorough treatment of the potential impact of varying machine learning methods for modelling, engineering and mitigation to address the problem \citep{rolnick2019tackling}. This review addresses the development of applied statistics and machine learning in the wider discipline of geoscience in the past 70 years and aims to provide context for the recent increase in interest and successes in machine learning and its challenges\footnote{The author of this manuscript has a background in geophysics, exploration geoscience, and active source 4D seismic. While this skews the expertise, they attempt to give a full overview over developments in all of geoscience with the minimum amount of bias possible.}. 

Machine learning (ML) is deeply rooted in applied statistics, building computational models that use inference and pattern recognition instead of explicit sets of rules. Machine learning is generally regarded as a sub-field of artificial intelligence (AI), with the notion of AI first being introduced by \citet{turing1950}. \citet{samuel1959some} coined the term machine learning itself, with \citet{mitchell1997machine} providing a commonly quoted definition: 

\begin{quote}
A computer program is said to learn from experience E with respect to some class of tasks T and performance measure P if its performance at tasks in T, as measured by P, improves with experience E. \begin{flushright}\citet{mitchell1997machine}\end{flushright}
\end{quote}

This means that a machine learning model is defined by a combination of requirements. A task such as, classification, regression, or clustering is improved by conditioning of the model on a training data set. The performance of the model is measured with regard to a loss, also called metric, which quantifies the performance of a machine learning model on the provided data. In regression, this would be measuring the misfit of the data from the expected values. Commonly, the model improves with exposure to additional samples of data. Eventually, a good model generalizes to unseen data, which was not part of the training set, on the same task the model was trained to perform.

Accordingly, many mathematical and statistical methods and concepts, including Bayes' rule \citep{bayes1763lii}, least-squares \citep{legendre1805nouvelles}, and Markov models \citep{markov1906rasprostranenie,markov1971extension}, are applied in machine learning. Gaussian processes stand out as they originate in time series applications \citep{kolmogorov1939interpolation} and geostatistics \citep{Krige1951}, which roots this machine learning application in geoscience \citep{rasmussen2003gaussian}. "Kriging" originally applied two-dimensional Gaussian processes to the prediction of gold mine valuation and has since found wide application in geostatistics. Generally, \citet{matheron1963principles} is credited with formalizing the mathematics of kriging and developing it further in the following decades.

Between 1950 and 2020 much has changed. Computational resources are now widely available both as hardware and software, with high-performance compute being affordable to anyone from cloud computing vendors. High-quality software for machine learning is widely available through the free and open-source software movement, with major companies (Google, Facebook, Microsoft) competing for the usage of their open-source machine learning frameworks (Tensorflow, Pytorch, CNTK\footnote{Deprecated 2019}) and independent developments reaching wide applications such as scikit-learn \citep{scikit-learn} and xgboost \citep{xgboost}. 

\begin{figure}
    \centering
    
    \includegraphics[width=\textwidth]{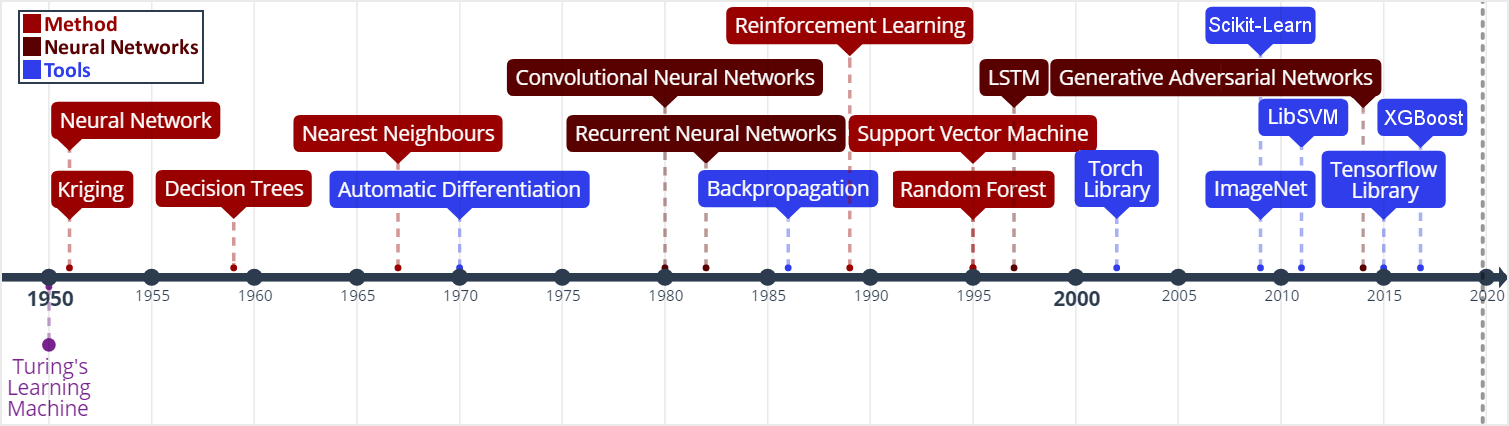}
    \caption{Machine Learning timeline from \citep{dramschphd}. Neural Networks: \citep{russelnorvig}; Kriging: \citep{Krige1951}; Decision Trees: \citep{belson1959matching}; Nearest Neighbours: \citep{cover1967nearest}; Automatic Differentiation: \citep{linnainmaa1970representation}; Convolutional Neural Networks: \citep{fukushima1980neocognitron, lecun2015deep}; Recurrent Neural Networks: \citep{hopfield1982neural}; Backpropagation: \citep{kelley1960gradient,bryson1961gradient,dreyfus1962numerical,rumelhart1988learning}; Reinforcement Learning: \citep{watkins1989learning}; Support Vector Machines: \citep{cortes1995support}; Random Forests: \citep{ho1995random}; LSTM: \citep{hochreiter1997long}; Torch Library: \citep{collobert2002torch}; ImageNet: \citep{deng2009imagenet}; Scikit-Learn: \citep{scikit-learn}; LibSVM: \citep{libsvm}; Generative Adversarial Networks: \citep{goodfellow2014generative}; Tensorflow: \citep{tensorflow}; XGBoost: \citep{xgboost}}
    \label{fig:ml-timeline}
\end{figure}

Nevertheless, investigations of machine learning in geoscience are not a novel development. The research into machine learning follows interest in artificial intelligence closely. Since its inception, artificial intelligence has experienced two periods of a decline in interest and trust, which has impacted negatively upon its funding. Developments in geoscience follow this wide-spread cycle of enthusiasm and loss of interest with a time lag of a few years. This may be the result of a variety of factors, including research funding availability and a change in willingness to publish results.

\section{Historic Machine Learning in Geoscience}
The 1950s and 1960s were decades of machine learning optimism, with machines learning to play simple games and perform tasks like route mapping. Intuitive methods like k-means, Markov models, and decision trees have been used as early as the 1960s in geoscience. K-means was used to describe the cyclicity of sediment deposits \citep{preston1964fourier}. \citet{krumbein1969markov} give a thorough treatment of the mathematical foundations of Markov chains and embedded Markov chains in a geological context through application to sedimentological processes, which also provides a comprehensive bibliography of Markov processes in geology. Some selected examples of early applications of Markov chains are found in sedimentology \citep{schwarzacher1972semi}, well log analysis \citep{agterberg1966markov}, hydrology \citep{matalas1967mathematical}, and volcanology \citep{wickman1968repose}. Decision tree-based methods found early applications in economic geology and prospectivity mapping \citep{newendorp1976decision,reddy1991decisiontree}. 

The 1970s were left with few developments in both the methods of machine learning, as well as, applications and adoption in geoscience (cf. Figure \ref{fig:ml-timeline}), due to the "first AI winter" after initial expectations were not met. Nevertheless, as kriging was not considered an AI technology, it was unaffected by this cultural shift and found applications in mining \citep{huijbregts1970universal}, oceanography \citep{chiles1975kriging}, and hydrology \citep{delhomme1978kriging}. This was in part due to superior results over other interpolation techniques, but also the provision of uncertainty measures. 

\subsection{Expert Systems to Knowledge-Driven AI}
The 1980s marked uptake in interest in machine learning and artificial intelligence through so-called "expert systems" and corresponding specialized hardware. While neural networks were introduced in 1950, the tools of automatic differentiation and backpropagation for error-correcting machine learning were necessary to spark their adoption in geophysics in the late 1980s. \citet{Zhao1988-hu} performed seismic deconvolution with a recurrent neural network (Hopfield network). \citet{Dowla1990-rd} discriminated between natural earthquakes and underground nuclear explosions using feed-forward neural networks. An ensemble of networks was able to achieve 97~\% accuracy for nuclear monitoring. Moreover, the researchers inspected the network to gain the insight that the ratio of particular input spectra was beneficial to the discrimination of seismological events to the network. However, in practice the neural networks underperformed on uncurated data, which is often the case in comparison to published results. \citet{Huang1990-hj} presented work on self-organizing maps (also Kohonen networks), a special type of unsupervised neural network applied to pick seismic horizons. The field of geostatistics saw a formalization of theory and an uptake in interest with \citet{matheron1981splines} formalizing the relationship of spline-interpolation and kriging and \citet{dubrule1984comparing} further develop the theory and apply it to well data. At this point, kriging is well-established in the mining industry as well as other disciplines that rely on spatial data, including the successful analysis and construction of the Channel tunnel \citep{chiles2018fifty}. The late 1980s then marked the second AI winter, where expensive machines tuned to run "expert systems" were outperformed by desktop hardware from non-specialist vendors, causing the collapse of a half-billion-dollar hardware industry. Moreover, government agencies cut funding in AI specifically. 

The 1990s are generally regarded as the shift from a knowledge-driven to a data-driven approach in machine learning. The term AI and especially expert systems were almost exclusively used in computer gaming and regarded with cynicism and as a failure in the scientific world. In the background, however, with research into applied statistics and machine learning, this decade marked the inception of Support-Vector Machines (SVM) \citep{cortes1995support}, the tree-based method Random Forests (RF) \citep{ho1995random}, and a specific type of recurrent neural network (RNN) Long Short-Term Memories (LSTM) \citep{hochreiter1997long}. SVMs were utilized for land usage classification in remote sensing early on \citep{hermes1999support}. Geophysics applied SVMs a few years later to approximate the Zoeppritz equations for AVO inversion, outperforming linearized inversion \citep{kuzma2003support}. Random Forests, however, were delayed in broader adoption, due to the term "random forests" only being coined in 2001 \citep{breiman2001random} and the statistical basis initially being less rigorous and implementation being more complicated. LSTMs necessitate large amounts of data for training and can be expensive to train, after further development in 2011 \citep{ciresan2011flexible} it gained popularity in commercial time series applications particularly speech and audio analysis.

\subsection{Neural Networks}
\citet{McCormack1991-pm} marks the first review of the emerging tool of neural networks in geophysics. The paper goes into the mathematical details and explores pattern recognition. The author summarizes neural network applications over the 30 years prior to the review and presents worked examples in automated well-log analysis and seismic trace editing. The review comes to the conclusion that neural networks are, in fact, good function approximators, taking over tasks that were previously reserved for human work. He criticizes slow training, the cost of retraining networks upon new knowledge, imprecision of outputs, non-optimal training results, and the black box property of neural networks. The main conclusion sees the implementation of neural networks in conventional computation and expert systems to leverage the pattern recognition of networks with the advantages of conventional computer systems.

Neural networks are the primary subject of the modern day machine learning interest, however, significant developments leading up to these successes were made prior to the 1990s. The first neural network machine was constructed by Minsky [described in \citet{russelnorvig}] and soon followed by the "Perceptron", a binary decision boundary learner \citep{rosenblatt1958perceptron}. This decision was calculated as follows:


\begin{equation}
\begin{array}{ll}
    {\color{DesatCyan}o_{j}} & = \sigma \left({\color{DesatMagenta}\sum_j w_{ij} x_{i} + b}\right)\\
    & = \sigma \left({\color{DesatMagenta}a_j}\right)\\
    & = \begin{cases}1&{\color{DesatMagenta}a_j} > 0 \\
    0 &\text{otherwise}   \end{cases}
\end{array}
\label{eq:perceptron}
\end{equation}
It describes a linear system with the output $o$, the linear activation $a$ of the input data $x$, the index of the source $i$ and target node $j$, the trainable weights $w$, the trainable bias $b$ and a binary activation function $\sigma$. The activation function $\sigma$ in particular has received ample attention since its inception. During this period, a binary $\sigma$ became uncommon and was replaced by non-linear mathematical functions. Neural networks are commonly trained by gradient descent, therefore, differentiable functions like sigmoid or tanh, allowing for the activation ${\color{DesatCyan}o}$ of each neuron in a neural network to be continuous. 

\begin{figure}
    \centering
    \includegraphics[width=.7\textwidth]{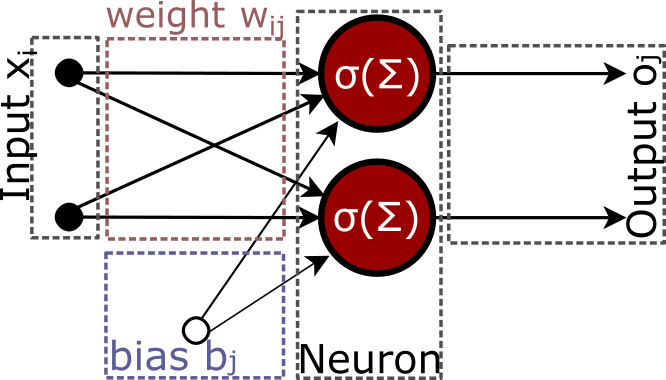}
    \caption{Single layer neural network as described in equation~\ref{eq:perceptron}. Two inputs $x_i$ are multiplied by the weights $w_{ij}$ and summed with the biases $b_j$. Subsequently an activation function $\sigma$ is applied to obtain out outputs $o_j$.}
    \label{fig:shallownn}
\end{figure}

Deep learning \citep{dechter1986learning} expands on this concept. It is the combination of multiple layers of neurons in a neural network. These deep networks learn representations with multiple levels of abstraction and can be expressed using equation~\ref{eq:perceptron} as {\color{DesatCyan}input neurons} to the next layer

\begin{equation}
\begin{array}{ll}
    o_k & = \sigma \left(\sum_k w_{jk} \cdot {\color{DesatCyan}o_j} + b\right)\\
    & = \sigma \left(\sum_k w_{jk} \cdot {\color{DesatCyan} \sigma \left(\sum_j w_{ij} x_{i} + b\right)} + b\right)
\end{array}
\label{eq:deepnetwork}
\end{equation}

\begin{figure}
    \centering
    \includegraphics[width=\textwidth]{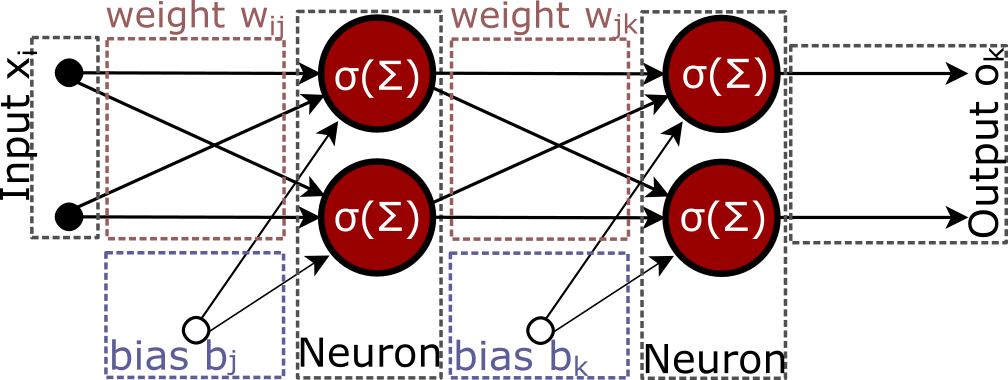}
    \caption{Deep multi-layer neural network as described in equation~\ref{eq:deepnetwork}.}
    \label{fig:deepnn}
\end{figure}

\citet{Roth1994-na} apply these building blocks of multi-layered neural networks with sigmoid activation to perform seismic inversion. They successfully invert low-noise and noise-free data on small training data. The authors note that the approach is susceptible to errors at low signal-to-noise ratios and coherent noise sources. Further applications include electromagnetic subsurface localization \citep{Poulton1992-ft}, magnetotelluric inversion via Hopfield neural networks \citep{Zhang1997-yp}, and geomechanical microfractures modelling in triaxial compression tests \citep{Feng1998-ck}.

\begin{figure}[H]
    \centering
    \includegraphics[width=\textwidth]{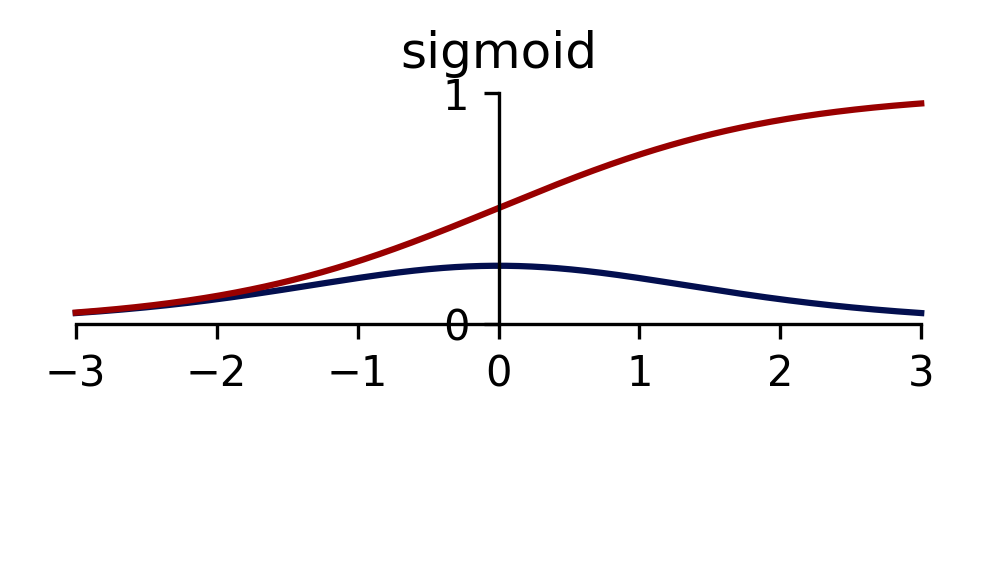}
    \caption{Sigmoid activation function (red) and derivative (blue) to train multi-layer Neural Network described in equation~\ref{eq:deepnetwork}.}
    \label{fig:mlp}
\end{figure}

\subsection{Kriging and Gaussian Processes}

\citet{cressie1990origins} review the history of kriging, prompted by the uptake of interest in geostatistics. The author defines kriging as Best Linear Unbiased Prediction and reviews the historical co-development of disciplines. Similar concepts were developed with mining, meteorology, physics, plant and animal breeding, and geodesy that relied on optimal spatial prediction. Later, \citet{williams1998} provide a thorough treatment of Gaussian Processes, in the light of recent successes of neural networks. 

\begin{quote}
An alternative method of putting a prior over functions is to use a Gaussian process (GP) prior over functions. This idea has been used for a long time in the spatial statistics community under the name of "kriging", although it seems to have been largely ignored as a general-purpose regression method. \begin{flushright}\citet{williams1998}\end{flushright}
\end{quote}

Overall, Gaussian Processes benefit from the fact that a Gaussian distribution will stay Gaussian under conditioning. That means that we can use Gaussian distributions in this machine learning process and they will produce a smooth Gaussian result after conditioning on the training data. To become a universal machine learning model, Gaussian Processes have to be able to describe infinitely many dimensions. Instead of storing infinite values to describe this random process, Gaussian Processes go the path of describing a distribution over functions that can produce each value when required. 

\begin{equation}
    p(x)\approx\mathcal{GP}\left(\mu(x),k(x, x')\right),
\end{equation}

The multivariate distribution over functions $p(x)$ is described by the Gaussian Process depends on mean a function $\mu(x)$ and a covariance function $k(x, x')$. It follows that choosing an appropriate mean and covariance function, also known as kernel, is essential. Very commonly, the mean function is chosen to be zero, as this simplifies some of the math. Therefore, data with a non-zero mean is commonly centered to comply with this assumption \citep{goertler2019a}. Choosing an appropriate kernel for the machine learning task is one of the benefits of the Gaussian Process. The kernel is where expert knowledge can be incorporated into data, e.g. seasonality metereological data can be described by a periodic covariance function.

\begin{figure}
    \centering
    
    \includegraphics[width=\textwidth]{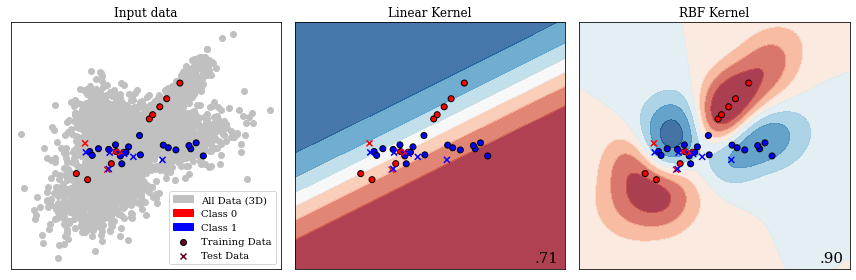}
    \caption{Gaussian Process separating two classes with different kernels. This image presents a 2D slice out of a 3D decision space. The decision boundary learnt from the data is visible, as well as the prediction in every location of the 2D slice. The two kernels presented are a linear kernel and a radial basis function (RBF) kernel, which show a significant discrepancy in performance. The bottom right number shows the accuracy on unseen test data. The linear kernel achieves $71~\%$ accuracy, while the RBF kernel achieves $90~\%$.}
    \label{fig:gp}
\end{figure}

Figure~\ref{fig:gp} present a 2D slice of 3D data with two classes. This binary problem can be approached by applying a Gaussian Process to it. In the second panel, a linear kernel is shown, which predicts the data relatively poorly with an accuracy of $71~\%$. A radial basis function (RBF) kernel, shown in the third panel generalizes to unseen test data with an accuracy of $90~\%$. This figure shows how a trained Gaussian Process would predict any new data point presented to the model. The linear kernel would predict any data in the top part to be blue (Class 0) and any data in the bottom part to be red (Class 1). The RBF kernel, which we explore further in the section introducing support-vector machines, separates the prediction into four uneven quadrants. The choice of kernel is very important in Gaussian Processes and research into extracting specific kernels is ongoing \citep{duvenaud2014automatic}.

In a more practical sense, Gaussian processes are computationally expensive, as an $n\times n$ matrix must be inverted, with $n$ being the number of samples. This results in a space complexity of $\mathcal{O}(n^2)$ and a time complexity $\mathcal{O}(n^3)$ \citep{williams2006gaussian}. This makes Gaussian Processes most feasible for smaller data problems, which is one explanation for their rapid uptake in geoscience. An approximate computation of the inverted matrix is possible using the Conjugate Gradient (CG) optimization method, which can be stopped early with a maximum  time cost of $\mathcal{O}(n^3)$ \citep{williams2006gaussian}. For problems with larger data sets, neural networks become feasible due to being computationally cheaper than Gaussian Processes, regularization on large data sets being viable, as well as, their flexibility to model a wide variety of functions and objectives. Regularization being essential as neural networks tend to not "overfit" and simply memorize the training data, instead of learning a generalizable relationship of the data. Interestingly, \citet{Hornik1989-bl} showed that neural networks are a universal function approximator as the number of weights tend to infinity, and \citet{neal1996} were able to show that the infinitely wide stochastic neural network converges to a Gaussian Process. Oftentimes Gaussian Processes are trained on a subset of a large data set to avoid the computational cost. Gaussian Processes have seen successful application on a wide variety of problems and domains that benefit from expert knowledge.

The 2000s were opened with a review by \citet{Van_der_Baan2000-jz} recapitulating the most recent geophysical applications in neural networks. They went into much detail on the neural networks theory and the difficulties in building and training these models. The authors identify the following subsurface geoscience applications through history: First-break picking, electromagnetics, magnetotellurics, seismic inversion, shear-wave splitting, well log analysis, trace editing, seismic deconvolution, and event classification. They reveal a strong focus on exploration geophysics. The authors evaluated the application of neural networks as subpar to physics-based approaches and concluded that neural networks are too expensive and complex to be of real value in geoscience. This sentiment is consistent with the broader perception of artificial intelligence during this decade. Artificial intelligence and expert systems over-promised human-like performance, causing a shift in focus on research into specialized sub-fields, e.g. machine learning, fuzzy logic, and cognitive systems. 

\section{Contemporary Machine Learning in Geoscience}

\citet{Mjolsness2001-fq} review machine learning in a broader context outside of exploration geoscience. The authors discuss recent successes in applications of remote sensing and robotic geology using machine learning models. They review graphical models, (hidden) Markov models, and SVMs and go on to disseminate the limitations of applications to vector data and poor performance when applied to rich data, such as graphs and text data. Moreover, the authors from NASA JPL go into detail on pattern recognition in automated rovers to identify geological prospects on Mars. They state:
\begin{quote}
The scientific need for geological feature catalogs has led to multiyear human surveys of Mars orbital imagery yielding tens of thousands of cataloged, characterized features including impact craters, faults, and ridges. \begin{flushright}\citet{Mjolsness2001-fq}\end{flushright}
\end{quote}
The review points out the profound impact SVMs have on identifying geomorphological features without modelling the underlying processes.

\subsection{Modern  Machine Learning Tools}
This decade of the 2000s introduces a shift in tooling, which is a direct contributor to the recent increase in adoption and research of both shallow and deep machine learning research.

Machine Learning software has been primarily comprised of proprietary software like Matlab\texttrademark with the Neural Networks Toolbox and Wolfram Mathematica\texttrademark or independent university projects like the Stuttgart Neural Network Simulator (SNNS). These tools were generally closed source and hard or impossible to extend and could be difficult to operate due to limited accompanying documentation. Early open-source projects include WEKA \citep{witten2005practical}, a graphical user interface to build machine learning and data mining projects. Shortly after that, LibSVM was released as free open-source software (FOSS) \citep{libsvm}, which implements support vector machines efficiently. It is still used in many other libraries to this day, including WEKA \citep{libsvm}. Torch was then released in 2002, which is a machine learning library with a focus on neural networks. While it has been discontinued in its original implementation in the programming language Lua \citep{collobert2002torch}, PyTorch, the reimplementation in the programming language Python, is one of the leading deep learning frameworks at the time of writing \citep{pytorch}. In 2007, the libraries Theano and scikit-learn were released openly licensed in Python \citep{theano, scikit-learn}. Theano is a neural network library that was a tool developed at the Montreal Institute for Learning Algorithms (MILA) and ceased development in 2017 after strong industrial developers had released openly licensed deep learning frameworks. Scikit-learn implements many different machine learning algorithms, including SVMs, Random Forests and single-layer neural networks, as well as utility functions including cross-validation, stratification, metrics and train-test splitting, necessary for robust machine learning model building and evaluation. 

\subsection{Support-Vector Machines}
\label{ssec:svm}
The impact of scikit-learn has shaped the current machine learning software package by implementing a unified application programming interface (API) \citep{sklearn_api}. This API is explored by example in the following code snippets, the code can be obtained at \citet{dramsch_2020_code_70}. First, we generate a classification dataset using a utility function. The \texttt{make\_classification} function takes different arguments to adjust the desired arguments, we are generating 5000 samples (\texttt{n\_samples}) for two classes, with five features (\texttt{n\_features}), of which three features are actually relevant to the classification (\texttt{n\_informative}). The data is stored in $X$, whereas the labels are contained in $y$.
\begin{minted}[mathescape,frame=lines,framesep=2mm]{python}
# Generate random classification dataset for example
from sklearn.datasets import make_classification
X, y = make_classification(n_samples=5000, n_features=5,
                           n_informative=3, n_redundant=0,
                           random_state=0, shuffle=False)
\end{minted}
It is good practice to divide the available labeled data into a training data set and a validation or test data set. This split ensures that models can be evaluated on unseen data to test the generalization to unseen samples. The utility function \texttt{train\_test\_split} takes an arbitrary amount of input arrays and separates them according to specified arguments. In this case 25\% of the data are kept for the hold-out validation set and not used in training. The \texttt{random\_state} is fixed to make these examples reproducible.
\begin{minted}[mathescape,frame=lines,framesep=2mm]{python}
# Split data into train and validation set
from sklearn.model_selection import train_test_split
X_train, X_test, y_train, y_test = train_test_split(X, y, 
                                            test_size=.25, 
                                            random_state=0)
\end{minted}
Then we need to define a machine learning model, considering the previous discussion of high impact machine learning models, the first example is an SVM classifier. This example uses the default values for hyperparameters of the SVM classifier, for best results on real-world problems these have to be adjusted. The machine learning training is always done by calling \texttt{classifier.fit(X, y)} on the classifier object, which in this case is the SVM object. In more detail, the \texttt{.fit()} method implements an optimization loop that will condition the model to the training data by minimizing the defined loss function. In the case of the SVM classification the parameters are adjusted to optimize a hinge loss, outlined in equation~\ref{eq:hingeloss}. The trained model scikit-learn model contains information about all its hyperparameters in addition to the trained model, shown below. The exact meaning of all these hyperparameters is laid out in the scikit-learn documentation \citep{sklearn_api}.

\begin{minted}[mathescape,frame=lines,framesep=2mm]{python}
# Define and train a Support Vector Machine Classifier
from sklearn.svm import SVC
svm = SVC(random_state=0)
svm.fit(X_train, y_train)

>>> SVC(C=1.0, break_ties=False, cache_size=200, 
        class_weight=None, coef0=0.0, degree=3, 
        decision_function_shape='ovr', gamma='scale', 
        kernel='rbf', max_iter=-1, probability=False, 
        random_state=0, shrinking=True, tol=0.001, 
        verbose=False)
\end{minted}
The trained SVM can the be used to predict on new data, by calling \texttt{classifier.predict(data)} on the trained classifier object. The new data has to contain four features like the training data did. Generally, machine learning models always need to be trained on the same set of input features as the data available for prediction. The \texttt{.predict()} method outputs the most likely estimate on the new data to generate predictions. In the following code snippet, three predictions on three input vectors are performed on the previously trained model.
\begin{minted}[mathescape,frame=lines,framesep=2mm]{python}
# Predict on new data with trained SVM
print(svm.predict([[0, 0, 0, 0, 0], 
                  [-1, -1, -1, -1, -1], 
                  [1, 1, 1, 1, 1]]))
>>> [1 0 1]
\end{minted}
The blackbox model should be evaluated with the \texttt{classifier.score()} function. Evaluating the performance on the training data set gives an indication how well the model is performing, but this is generally not enough to gauge the performance of machine learning models. In addition, the trained model has to be evaluated on the hold-out set, a dataset the model has not been exposed to during training. This avoids that the model only performs well on the training data by "memorization" instead of extracting meaningful generalizable relationships, an effect called overfitting. In this example the hyperparameters are left to the default values, in real-life applications hyperparameters are usually adjusted to build better models. This can lead to an addition meta-level of overfitting on the hold-out set, which necessitates an additional third hold-out set to test the generalizability of the trained model with optimized hyperparameters. The default score uses the class accuracy, which suggests our model is approximately 90\% correct. Similar train and test scores indicate that the model learned a generalizable model, enabling prediction on unseen data without a performance loss. Large differences between the training score and test score indicate either overfitting, in the case of a better training score. A higher test score than training score can be an indication of a deeper problem with the data split, scoring, class imbalances, and needs to be investigated by means of external cross-validation, building standard "dummy" models, independence tests, and further manual investigations.
\begin{minted}[mathescape,frame=lines,framesep=2mm]{python}
# Score SVM on train and test data
print(svm.score(X_train, y_train))
print(svm.score(X_test, y_test))
>>> 0.9098666666666667
>>> 0.9032
\end{minted}

\begin{figure}
    \centering
    \includegraphics[width=\textwidth]{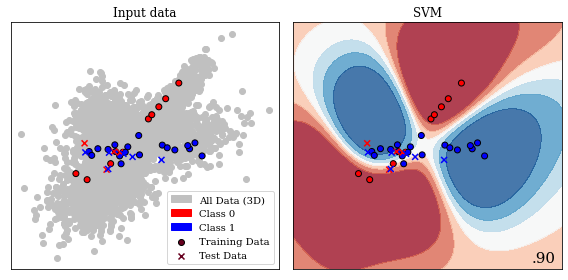}
    \caption{Example of Support Vector Machine separating two classes, showing the decision boundary learnt from the data. The data contains three informative features, the decision boundary is therefore three dimensional, shown is a central slice of data points in 2D. (A video is available at \citep{dramsch_2020_3d})}
    \label{fig:svm}
\end{figure}

Support-vector machines can be employed for each class of machine learning problem, i.e. classification, regression, and clustering. In a two-class problem, the algorithm considers the $n$-dimensional input and attempts to find a $(n-1)$-dimensional hyperplane that separates these input data points. The problem is trivial if the two classes are linearly separable, also called a hard margin. The plane can pass the two classes of data without ambiguity. For data with an overlap, which is usually the case, the problem becomes an optimization problem to fit the ideal hyperplane. The hinge loss provides the ideal loss function for this problem, yielding 0 if none of the data overlap, but a linear residual for overlapping points that can be minimized:
\begin{equation}
\max \left( 0, (1-y_i(\vec{w}\cdot \vec{x}_i - b)) \right),
\end{equation}
with $y_i$ being the current target label and $\vec{w}\cdot \vec{x}_i - b$ being the hyperplane under consideration. The hyperplane consists of $w$ the normal vector and point $x$, with the offset $b$. This leads the algorithm to optimize

\begin{equation}
\left[\frac 1 n \sum_{i=1}^n \max\left(0, 1 - y_i(w\cdot x_i - b)\right) \right] + \lambda\lVert w \rVert^2,
\label{eq:hingeloss}
\end{equation}

with $\lambda$ being a scaling factor. For small $\lambda$ the loss becomes the hard margin classifier for linearly separable problems. The nature of the algorithm dictates that only values for $\vec{x}$ close to the hyperplane define the hyperplane itself; these values are called the support vectors.

The SVM algorithm would not be as successful if it were simply a linear classifier. Some data can become linearly separable in higher dimensions. This, however, poses the question of how many dimensions should be searched, because of the exponential cost in computation that follows due to the increase of dimensionality (also known as the curse of dimensionality). Instead, the "kernel trick" was proposed \citep{aizerman}, which defines a set of values that are applied to the input data simply via the dot product. A common kernel is the radial basis function (RBF), which is also the kernel we applied in the example. The kernel is defined as:

\begin{equation}
k\left(\vec{x}_i, \vec{x}_j \right) \rightarrow \exp\left( -\gamma \lVert \vec{x}_i - \vec{x}_j \rVert^2 \right)
\end{equation}

\begin{figure}
    \centering
    \includegraphics[width=\textwidth]{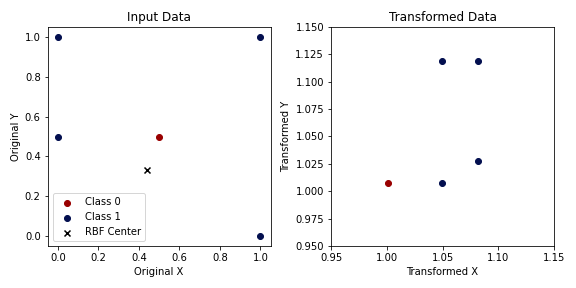}
    \caption{Samples from two classes that are not linearly separable input data (left).  Applying a Gaussian Radial Basis Function centered around $(0.4, 0.33)$ with $\lambda = .5$ results in the two classes being linearly separable.}
    \label{fig:rbf}
\end{figure}

This specifically defines the Gaussian Radial Basis Function of every input data point with regard to a central point. This transformation can be performed with other functions (or kernels), such as, polynomials or the sigmoid function. The RBF will transform the data according to the distance between $x_i$ and $X_j$, this can be seen in Figure~\ref{fig:rbf}. This results in the decision surface in Figure~\ref{fig:svm} consisting of various Gaussian areas. The RBF is generally regarded as a good default, in part, due to being translation invariant (i.e. stationary) and smoothly varying.

An important topic in machine learning is explainability, which inspects the influence of input variables on the prediction. We can employ the utility function \texttt{permutation\_importance} to inspect any model and how they perform with regard to their input features \citep{breiman2001random}. The permutation importance evaluates how well the blackbox model performs, when a feature is not available. Practically, a feature is replaced with random noise. Subsequently, the score is calculated, which provides a representation how informative a feature is compared to noise. The data we generated in the first example contains three informative features and two random data columns. The mean values of the calculated importances show that three features are estimated to be three magnitudes more important, with the second feature containing the maximum amount of information to predict the labels.

\begin{minted}[mathescape,frame=lines,framesep=2mm]{python}
# Calculate permutation importance of SVM model
from sklearn.inspection import permutation_importance
importances = permutation_importance(svm, X_train, y_train, 
                                     n_repeats=10, random_state=0)

# Show mean value of importances and the ranking
print(importances.importances_mean)
print(importances.importances_mean.argsort())
>>> [ 2.1787e-01  2.8712e-01  1.2293e-01 -1.8667e-04  7.7333e-04]
>>> [3 4 2 0 1]
\end{minted}

Support-vector machines were applied to seismic data analysis \citep{Li2004-fk} and the automatic seismic interpretation \citep{Liu2015-pf,di2017seismic,Mardan2017-vr}. Compared to convolutional neural networks, these approaches usually do not perform as well, when the CNN can gain information from adjacent samples. Seismological volcanic tremor classification \citep{Masotti2006-fi,Masotti2008-tu} and analysis of ground-penetrating radar \citep{pasolli2009automatic, Xie2013-fh} were other notable applications of SVM in Geoscience. The 2016 Society of Exploration Geophysicists (SEG) machine learning challenge was held using a SVM baseline \citep{Hall2016-xh}. Several other authors investigated well log analysis \citep{anifowose2017carbonate, Cate2018-mb, Gupta2018-ut, Saporetti2018-sq}, as well as seismology for event classification \citep{Malfante2018-yl} and magnitude determination \citep{Ochoa2018-wp}. These rely on SVMs being capable of regression on time-series data. Generally, many applications in geoscience have been enabled by the strong mathematical foundation of SVMs, such as microseismic event classification \citep{Zhao2017-rx}, seismic well ties \citep{Chaki2018-mr}, landslide susceptibility \citep{Marjanovic2011-ot,Ballabio2012-xv}, digital rock models \citep{Ma2012-qo}, and lithology mapping \citep{cracknell2013upside}.

\subsection{Random Forests}
The following example shows the application of Random Forests, to illustrate the similarity of the API for different machine learning algorithms in the scikit-learn library. The Random Forest classifier is instantiated with a maximum depth of seven, and the random state is fixed to zero again. Limiting the depth of the forest forces the random forest to conform to a simpler model. Random forests have the capability to become highly complex models that are very powerful predictive models. This is not conducive to this small example dataset, but easy to modify for the inclined reader. The classifier is then trained using the same API of all classifiers in scikit-learn. The example shows a very high number of hyperparameters, however, Random Forests work well without further optimization of these.
\begin{minted}[mathescape,frame=lines,framesep=2mm]{python}
# Define and train a Random Forest Classifier
from sklearn.ensemble import RandomForestClassifier
rf = RandomForestClassifier(max_depth=7, random_state=0)
rf.fit(X_train, y_train)

>>> RandomForestClassifier(bootstrap=True, ccp_alpha=0.0,
                class_weight=None, criterion='gini', max_depth=7, 
                max_features='auto', max_leaf_nodes=None, 
                max_samples=None, min_impurity_decrease=0.0, 
                min_impurity_split=None, min_samples_leaf=1, 
                min_samples_split=2, min_weight_fraction_leaf=0.0, 
                n_estimators=100, n_jobs=None, oob_score=False, 
                random_state=0, verbose=0, warm_start=False)
\end{minted}
The prediction of the random forest is performed in the same API call again, also consistent with all classifiers available. The values are slightly different from the prediction of the SVM.
\begin{minted}[mathescape,frame=lines,framesep=2mm]{python}
# Predict on new data with trained Random Forest
print(rf.predict([[0, 0, 0, 0, 0], 
                 [-1, -1, -1, -1, -1], 
                 [1, 1, 1, 1, 1]]))
>>> [1 0 1]
\end{minted}
The training score of the random forest model is 2.5~\% better than the SVM in this instance, this score however not informative. Comparing the test scores shows only a 0.88~\% difference, which is the relevant value to evaluate, as it shows the performance of a model on data it has not seen during the training stage. The random forest performed slightly better on the training set than the test data set. This slight discrepancy is usually not an indicator of an overfit model. Overfit models "memorize" the training data and do not generalize well, which results in poor performance on unseen data. Generally, overfitting is to be avoided in real application, but can be seen in competitions, on benchmarks, and show-cases of new algorithms and architectures to oversell the improvement over state-of-the-art methods \citep{recht2019ImageNet}.
\begin{minted}[mathescape,frame=lines,framesep=2mm]{python}
# Score Random Forest on train and test data
print(rf.score(X_train, y_train))
print(rf.score(X_test, y_test))
>>> 0.9306
>>> 0.912
\end{minted}
Random forests have specialized methods available for introspection, which can be used to calculate feature importance. These are based on the decision process the random forest used to build the machine learning model. The feature importance in Random Forests uses the same method as permutation importance, which is dropping out features to estimate their importance on the model performance. Random Forests use a measure to determine the split between classes at each node of the trees called Gini impurity. While the permutation importance uses the accuracy score of the prediction, in Random Forests this Gini impurity can be used to measure how informative a feature is in a model. It is important to note that this impurity-based process can be susceptible to noise and overestimate high number of classes in features. Using the permutation importance instead is a valid choice. In this instance as opposed to the permutation importance, the random forest estimates the two non-informative features to be one magnitude less useful than the informative features, instead of two magnitudes. 
\begin{minted}[mathescape,frame=lines,framesep=2mm]{python}
# Inspect random forest for feature importance
print(rf.feature_importances_)
print(rf.feature_importances_.argsort())
>>> [0.2324 0.4877 0.2527 0.0141 0.0129]
>>> [4 3 0 2 1]
\end{minted}

Random forests and other tree-based methods, including gradient boosting, a specialized version of random forests, have generally found wider application with the implementation into scikit-learn and packages for the statistical languages R and SPSS. Similar to neural networks, this method is applied to ASI \citep{Guillen2015-re} with limited success, which is due to the independent treatment of samples, like SVMs. Random forests have the ability to approximate regression problems and time series, which made them suitable for seismological applications including localization \citep{Dodge2016-ah}, event classification in volcanic tremors \citep{Maggi2017-mr} and slow slip analysis \citep{Hulbert2018-xe}. They have also been applied to geomechanical applications in fracture modelling \citep{Valera2017-yl} and fault failure prediction \citep{RouetLeduc2017-li,RouetLeduc2018-vd}, as well as, detection of reservoir property changes from 4D seismic data \citep{Cao2017-gp}. Gradient Boosted Trees were the winning models in the 2016 SEG machine learning challenge \citep{Hall2017-fk} for well-log analysis, propelling a variety of publications in facies prediction \citep{Bestagini2017-nh,Blouin2017-gt,Cate2018-mb,Saporetti2018-sq}.

\begin{figure}
    \centering
    \includegraphics[width=\textwidth]{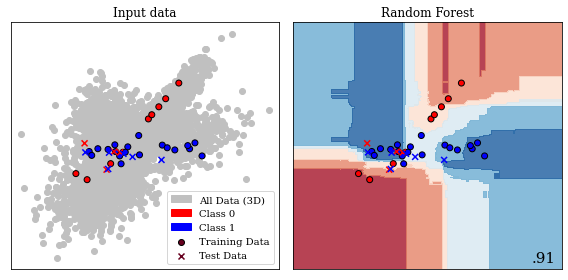}
    \caption{Binary Decision Boundary for Random Forest in 2D. This is the same central slice of the 3D decision volume used in Figure~\ref{fig:svm}.}
   \label{fig:randomforest}
\end{figure}

Furthermore, various methods that have been introduced into scikit-learn have been applied to a multitude of geoscience problems. Hidden Markov models were used on seismological event classification \citep{Ohrnberger2001-cq,Beyreuther2008-mz, Bicego2013-ox}, well-log classification \citep{Jeong2014-jy, Wang2017-gi}, and landslide detection from seismic monitoring \citep{Dammeier2016-mf}. These hidden Markov models are highly performant on time series and spatially coherent problems. The "hidden" part of Markov models enables the model to assume influences on the predictions that are not directly represented in the input data. The K-nearest neighbours method has been used for well-log analysis \citep{Cate2017-na,Saporetti2018-sq}, seismic well ties \citep{Wang2017-lw} combined with dynamic time warping and fault extraction in seismic interpretation \citep{hale2013methods}, which is highly dependent on choosing the right hyperparameter k. The unsupervised k-NN equivalent, k-means has been applied to seismic interpretation \citep{Di2017-qn}, ground motion model validation \citep{Khoshnevis2018-wq}, and seismic velocity picking \citep{Wei2018-nm}. These are very simple machine learning models that are useful for baseline models. Graphical modelling in the form of Bayesian networks has been applied to seismology in modelling earthquake parameters \citep{Kuehn2011-tv}, basin modelling \citep{Martinelli2013-ch}, seismic interpretation \citep{Ferreira2018-gr} and flow modelling in discrete fracture networks \citep{Karra2018-of}. These graphical models are effective in causal modelling and gained popularity in modern applications of machine learning explainability, interpretability, and generalization in combination with do-calculus \citep{pearl2012calculus}.

\subsection{Modern Deep Learning}
The 2010s marked a renaissance of deep learning and particularly convolutional neural networks. The convolutional neural network (CNN) architecture AlexNet \citep{krizhevsky2012ImageNet} was the first CNN to enter the ImageNet challenge \citep{deng2009imagenet}. The ImageNet challenge is considered a benchmark competition and database of natural images established in the field of computer vision. This improved the classification error rate from 25.8~\% to 16.4~\% (top-5 accuracy). This has propelled research in CNNs, resulting in error rates on ImageNet of 2.25~\% on top-5 accuracy in 2017 \citep{ImageNetresults}. The Tensorflow library \citep{tensorflow} was introduced for open source deep learning models, with some different software design compared to the Theano and Torch libraries.

The following example shows an application of deep learning to the data presented in the previous examples. The classification data set we use has independent samples, which leads to the use of simple densely connected feed-forward networks. Image data or spatially correlated datasets would ideally be fed to a convolutional neural network (CNN), whereas time series are often best approached with recurrent neural networks (RNN). This example is written using the Tensorflow library. PyTorch would be an equally good library to use.

All modern deep learning libraries take a modular approach to building deep neural networks that abstract operations into layers. These layers can be combined into input and output configurations in highly versatile and customizable ways. The simplest architecture, which is the one we implement below, is a sequential model, which consists of one input and one output layer, with a "stack" of layers. It is possible to define more complex models with multiple inputs and outputs, as well as the branching of layers to build very sophisticated neural network pipelines. These models are called functional API and subclassing API, but would not be conducive to this example.

The example model consists of Dense layers and a Dropout layer, which are arranged in sequence. Densely connected layers contain a specified number of neurons with an appropriate activation function, shown in the example below. Each neuron performs the calculation outlined in equation~\ref{eq:perceptron}, with $\sigma $ defining the activation. Modern neural networks rarely implement \texttt{sigmoid} and \texttt{tanh} activations anymore. Their activation characteristic leads them to lose information for large positive and negative values of the input, commonly called saturation\citep{hochreiter2001gradient}. This saturation of neurons prevented good deep neural network performance until new non-linear activation functions took their place\citep{xu2015empirical}. The activation function Rectified linear unit (ReLU) is generally credited with facilitating the development of very deep neural networks, due to their non-saturating properties \citep{hahnloser2000digital}. It sets all negative values to zero and provides a linear response for positive values, as seen in equation~\ref{eq:relu}. Since it's inception, many more rectifiers with different properties have been introduced.

\begin{equation}
    \sigma({\color{DesatMagenta} a}) = max(0, {\color{DesatMagenta}a})
    \label{eq:relu}
\end{equation}

\begin{figure}
    \centering
    \includegraphics[width=\textwidth]{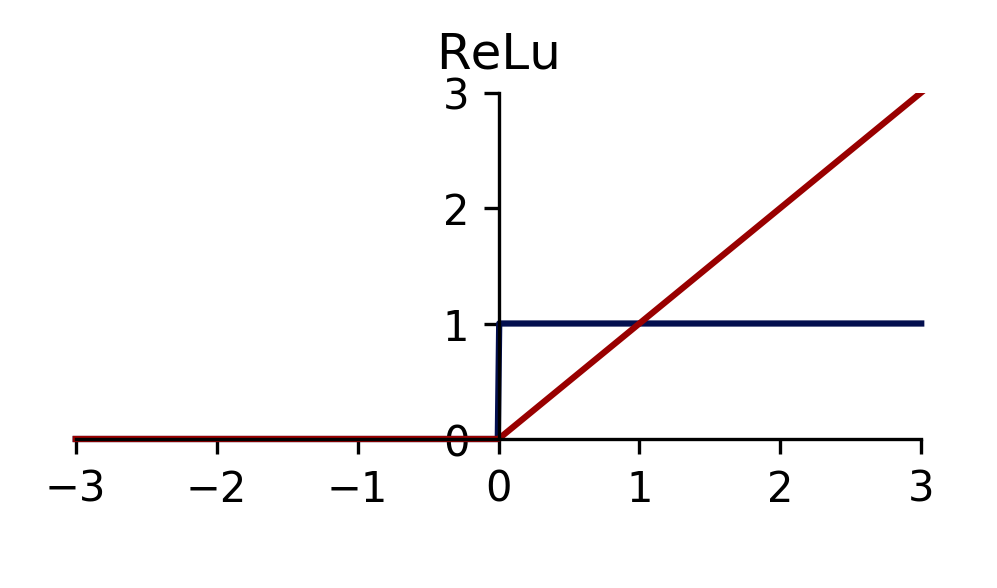}
    \caption{ReLU activation (red) and derivative (blue) for efficient gradient computation.}
    \label{fig:relu}
\end{figure}

The other activation function used in the example is the "softmax" function on the output layer. This activation is commonly used for classification tasks, as it normalizes all activations at all outputs to one. It achieves this by applying the exponential function to each of the outputs in ${\color{DesatMagenta}\vec{a}}$ for class $C$ and dividing that value by the sum of all exponentials:

\begin{equation}
\sigma({\color{DesatMagenta}\vec{a}}) = \frac{e^{{\color{DesatMagenta}a_j}}}{\sum\limits_{p}^C e^{{\color{DesatMagenta}a_p}}}
\label{eq:softmax}
\end{equation}

The example additionally uses a Dropout layer, which is a common layer used for regularization of the network by randomly setting a specified percentage of nodes to zero for each iteration. Neural networks are particularly prone to overfitting, which is counteracted by various regularization techniques that also include input-data augmentation, noise injection, $\mathcal{L}_1$ and $\mathcal{L}_2$ constraints, or early-stopping of the training loop \citep{deeplearningbook}. Modern deep learning systems may even leverage noisy student-teacher networks for regularization \citep{xie2019self}.

\begin{minted}[mathescape,frame=lines,framesep=2mm]{python}
import tensorflow as tf
model = tf.keras.models.Sequential([
tf.keras.layers.Dense(32, activation='relu'),
tf.keras.layers.Dropout(.3),
tf.keras.layers.Dense(16, activation='relu'),
tf.keras.layers.Dense(2, activation='softmax')])
\end{minted}

These sequential models are also used for simple image classification models using CNNs. Instead of Dense layers, these are built up with convolutional layers, which are readily available in 1D, 2D, and 3D as Conv1D, Conv2D and Conv3D respectively. A two-dimensional CNN learns a so-called filter $f$ for the $n\times m$-dimensional image $G$, expressed as:
\begin{equation}
G^{*}(x,y) = \sum_{i=1}^{n} \sum_{j=1}^{m} f(i,j)\cdot G(x-i+c,\; y-j+c),
\label{eq:convolution}
\end{equation}
resulting in the central result $G^{*}$ around the central coordinate $c$. In CNNs each layer learns several of these filters $f$, usually following by a down-sampling operation in $n$ and $m$ to compress the spatial information. This serves as a forcing function to learn increasingly abstract representations in subsequent convolutional layers. 

\begin{figure}
    \centering
    \includegraphics[width=.5\textwidth]{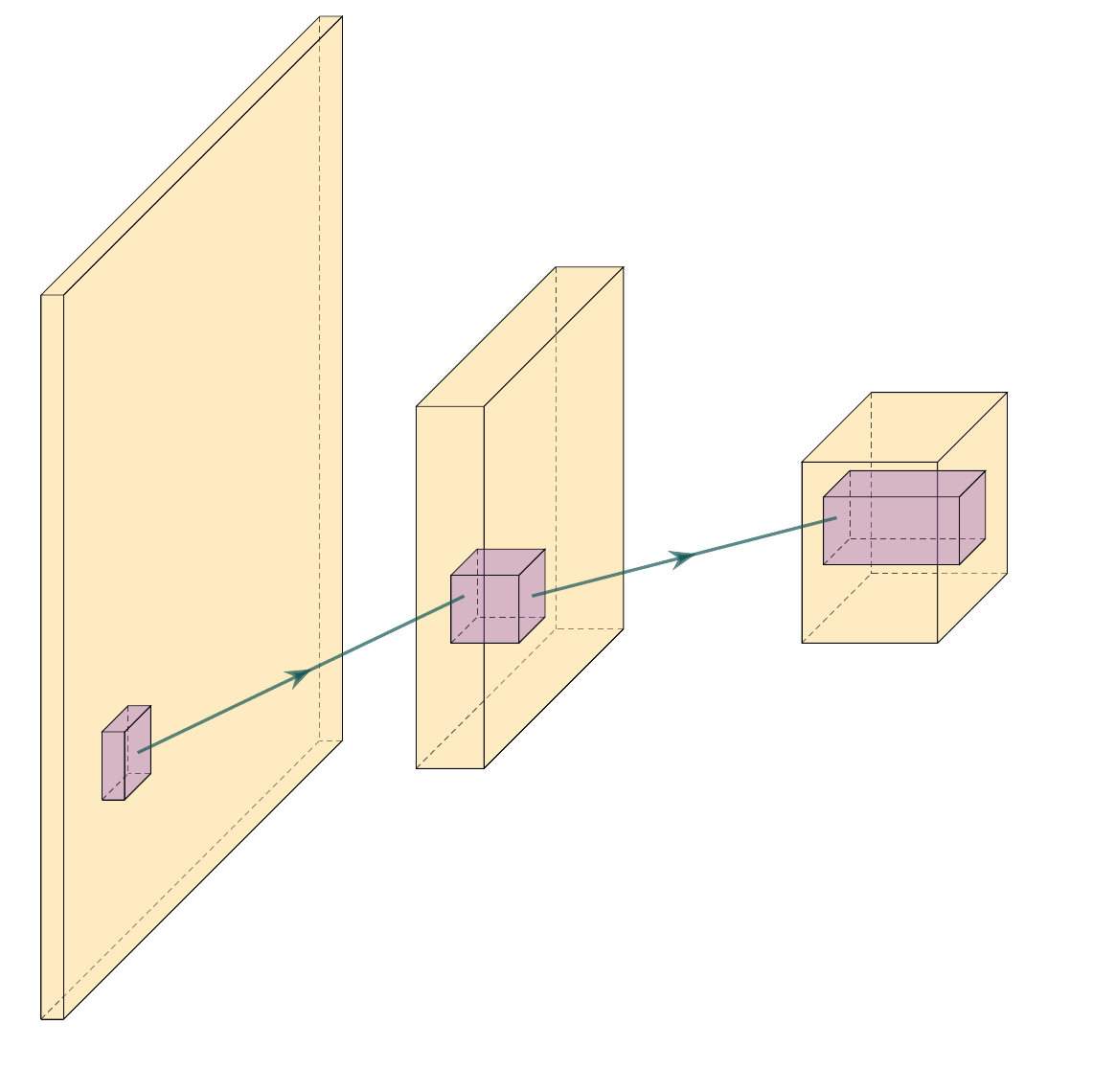}
    \caption{Three layer convolutional network. The input image (yellow) is convolved with several filters or kernel matrices (purple). Commonly, the convolution is used to downsample an image in the spatial dimension, while expanding the dimension of the filter response, hence expanding in "thickness" in the schematic. The filters are learned in the machine learning optimization loop. The shared weights within a filter improve efficiency of the network over classic dense networks.}
   \label{fig:cnn}
\end{figure}

This sequential example model of densely connected layers with a single input, 32, 16, and two neurons contains a total of 754 trainable weights. Initially, each of these weights is set to a pseudo-random value, which is often drawn from a distribution beneficial to fast training. Consequently, the data is passed through the network, and the result is numerically compared to the expected values. This form of training is defined as supervised training and error-correcting learning, which is a form of Hebbian learning. Other forms of learning exist and are employed in machine learning, e.g. competitive learning in self-organizing maps.

\begin{equation}
    MAE = \lvert{y_j - \color{DesatCyan}o_{j}}\rvert \newline
    \label{eq:mae}
\end{equation}
\begin{equation}
    MSE = ({y_j - \color{DesatCyan}o_{j}})^2
    \label{eq:mse}
\end{equation}

In regression problems the error is often calculated using the Mean Absolute Error (MAE) or Mean Squared Error (MSE), the $\mathcal{L}_1$ shown in equation~\ref{eq:mae} and the $\mathcal{L}_2$ norm shown in equation~\ref{eq:mse} respectively. Classification problems form a special type of problem that can leverage a different kind of loss called cross-entropy (CE). The cross-entropy is dependent on the true label $y$ and the prediction in the output layer.

\begin{equation}
    CE = - \sum\limits^C_j y_j \log{\left({\color{DesatCyan}o_{j}}\right)}
    \label{eq:crossentropy}
\end{equation}

Many machine learning data sets have one true label $y_{true} = 1$ for class $C_{j = true}$, leaving all other $y_j = 0$. This makes the sum over all labels obsolete. It is debatable how much binary labels reflects reality, but it simplifies equation~\ref{eq:crossentropy} to minimizing the (negative) logarithm of the neural network output ${\color{DesatCyan}o_{j}}$, also known as negative log-likelihood:

\begin{equation}
    CE = - \log{\left({\color{DesatCyan}o_{j}}\right)}
    \label{eq:binarycrossentropy}
\end{equation}

Technically, the data we generated is a binary classification problem, and this means we could use the sigmoid activation function in the last layer and optimize a binary CE. This can speed up computation, but in this example, an approach is shown that works for many other problems and can therefore be applied to the readers data.

\begin{minted}[mathescape,frame=lines,framesep=2mm]{python}
model.compile(optimizer='adam', # Often 'adam' or 'sgd' are good
              loss='sparse_categorical_crossentropy',
              metrics=['accuracy']) # Monitor other metrics
\end{minted}

Large neural networks can be extremely costly to train with significant developments in 2019/2020 reporting multi-billion parameter language models (Google, OpenAI) trained on massive hardware infrastructure for weeks with a single epoch taking several hours. This calls for validation on unseen data after every epoch of the training run. Therefore, neural networks, like all machine learning models, are commonly trained with two hold-out sets, a validation and a final test set. The validation set can be provided or be defined as a percentage of the training data, as shown below. In the example, 10\% of the training data are held out for validation after every epoch, reducing the training data set from 3750 to 3375 individual samples. 

\begin{minted}[mathescape,frame=lines,framesep=2mm]{python}
model.fit(X_train, 
          y_train, 
          validation_split=.1,
          epochs=100)
>>> [...]
    Epoch 100/100
    3375/3375 [==============================] - 0s 66us/sample
    loss: 0.1567 - accuracy: 0.9401 - 
    val_loss: 0.1731 - val_accuracy: 0.9359
\end{minted}

Neural networks are trained with variations of stochastic gradient descent (SGD), an incremental version of the classic steepest descent algorithm. We use the Adam optimizer, a variation of SGD that converges fast, but a full explanation would go beyond the scope of this chapter. The gist of the Adam optimizer is that it maintains a per-parameter learning rate of the first statistical moment (mean). This is beneficial for sparse problems and the second moment (uncentered variance), which is beneficial for noisy and non-stationary problems \citep{kingma2014adam}. The main alternative to Adam is SGD with Nesterov momentum \citep{pmlr-v28-sutskever13}, an optimization method that models conjugate gradient methods (CG) without the heavy computation that comes with the search in CG. SGD anecdotally finds a better optimal point for neural networks than Adam but converges much slower.

In addition to the loss value, we display the accuracy metric. While accuracy should not be the sole arbiter of model performance, it gives a reasonable initial estimate, how many samples are predicted correctly with a percentage between zero and one. As opposed to scikit-learn, deep learning models are compiled after their definition to make them fit for optimization on the available hardware. Then the neural network can be fit like the SVM and Random Forest models before, using the \texttt{X\_train} and \texttt{y\_train} data. In addition, a number of epochs can be provided to run, as well as other parameters that are left on default for the example. The amount of epochs defines how many cycles of optimization on the full training data set are performed. Conventional wisdom for neural network training is that it should always learn for more epochs than machine learning researchers estimate initially.

\begin{figure}[H]
    \centering
    
    \includegraphics[width=\textwidth]{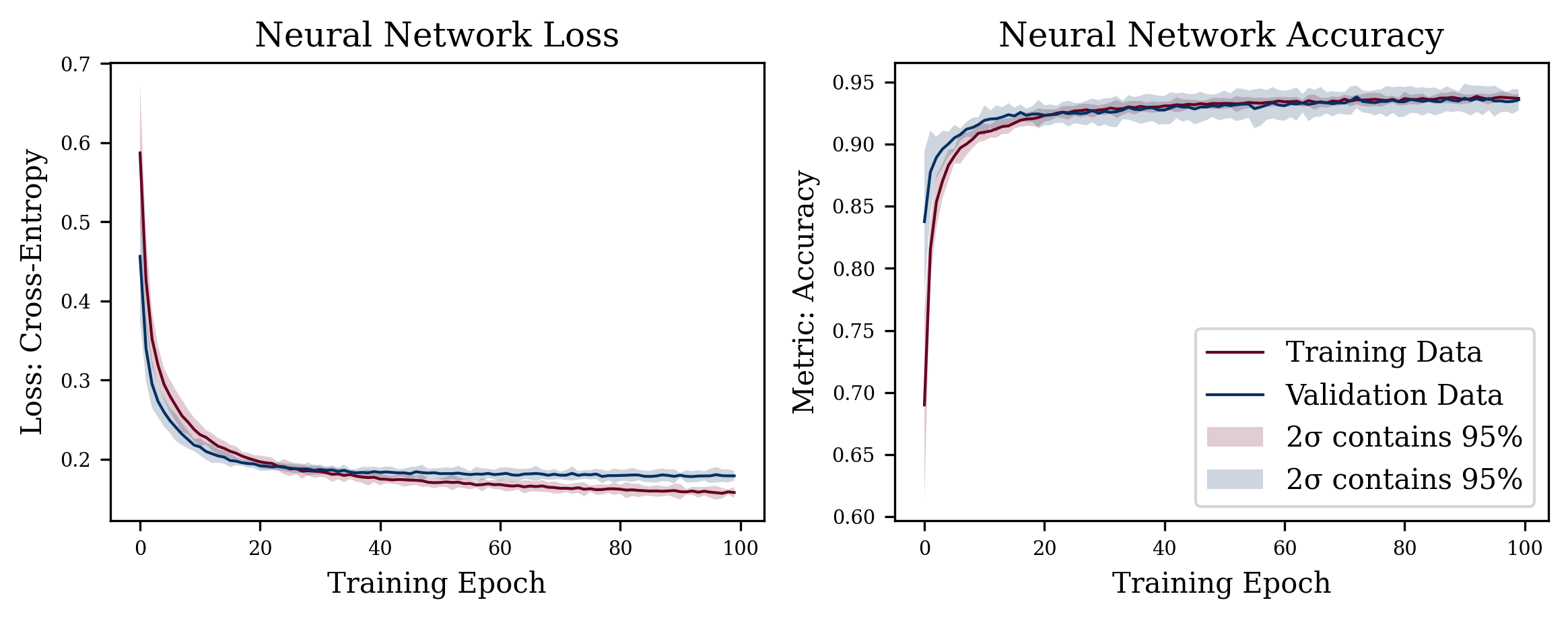}
    \caption{Loss and Accuracy of example neural network on ten random initializations. Training for 100 epochs with the shaded area showing the 95\% confidence intervals of the loss and metric. Analyzing loss curves is important to evaluate overfitting. The trining loss decreasing, while validation loss is close to plateauing is a sign of overfitting. Generally, it can be seen that the model converged and is only making marginal gains with the risk of overfitting.}
   \label{fig:training_loss}
\end{figure}

It can be difficult to fix all sources of randomness and stochasticity in neural networks, to make both research and examples reproducible. This example does not fix these so-called random seeds as it would detract from the example. That implies that the results for loss and accuracy will differ from the printed examples. In research fixing the seed is very important to ensure reproducibility of claims. Moreover, to avoid bad practices or so-called "lucky seeds", a statistical analysis of multiple fixed seeds is good practice to report results in any machine learning model.

\begin{minted}[mathescape,frame=lines,framesep=2mm]{python}
model.evaluate(X_test, y_test)
>>> 1250/1250 [==============================] - 0s 93us/sample
    loss: 0.1998 - accuracy: 0.9360
    [0.19976349686831235, 0.936]
\end{minted}

\begin{figure}
    \centering
    \includegraphics[width=\textwidth]{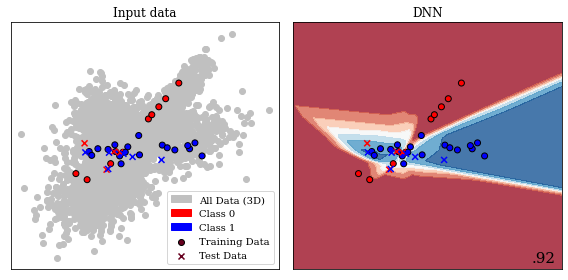}
    \caption{Central 2D slice of decision Boundary of deep neural network in trained on data with 3 informative features. The 3D volume is available in \citep{dramsch_2020_3d}.}
    \label{fig:dnndecision}
\end{figure}

In the example before, the SVM and Random Forest classifier were scored on unseen data. This is equally important for neural networks. Neural networks are prone to overfit, which we try to circumvent by regularizing the weights and by evaluating the final network on an unseen test set. The prediction on the test set is very close to the last epoch in the training loop, which is a good indicator that this neural network generalizes to unseen data. Moreover, the loss curves in figure~\ref{fig:training_loss} do not converge too fast, while converging. However, it appears that the network would overfit if we let training continue. The exemplary decision boundary in figure~\ref{fig:dnndecision} very closely models the local distribution of the data, which is true for the entire decision volume \citep{dramsch_2020_3d}.

These examples illustrate the open source revolution in machine learning software. The consolidated API and utility functions make it seem trivial to apply various machine learning algorithms to scientific data. This can be seen in the recent explosion of publications of applied machine learning in geoscience. The need to be able to implement algorithms has been replaced by merely installing a package and calling \texttt{model.fit(X, y)}. These developments call for strong validation requirements of models to ensure valid, reproducible, and scientific results. Without this careful validation these modern day tools can be severely misused to oversell results and even come to incorrect conclusions.

In aggregate, modern-day neural networks benefit from the development of non-saturating non-linear activation functions. The advancements of stochastic gradient descent with Nesterov momentum and the Adam optimizer (following AdaGrad and RMSProp) was essential faster training of deep neural networks. The leverage of graphics hardware available in most high-end desktop computers that is specialized for linear algebra computation, further reduced training times. Finally, open-source software that is well-maintained, tested, and documented with a consistent API made both shallow and deep machine learning accessible to non-experts.

\subsection{Neural Network Architectures}
In deep learning, implementation of models is commonly more complicated than understanding the underlying algorithm. Modern deep learning makes use of various recent developments that can be beneficial to the data set it is applied to, without specific implementation details results are often not reproducible. However, the machine learning community has a firm grounding in openness and sharing, which is seen in both publications and code. New developments are commonly published alongside their open-source code, and frequently with the trained networks on standard benchmark data sets. This facilitates thorough inspection and transferring the new insights to applied tasks such as geoscience. In the following, some relevant neural network architectures and their application are explored.

\subsection{Convolutional Neural Network Architectures}
\begin{figure}
    \centering
    
    \includegraphics[width=\textwidth]{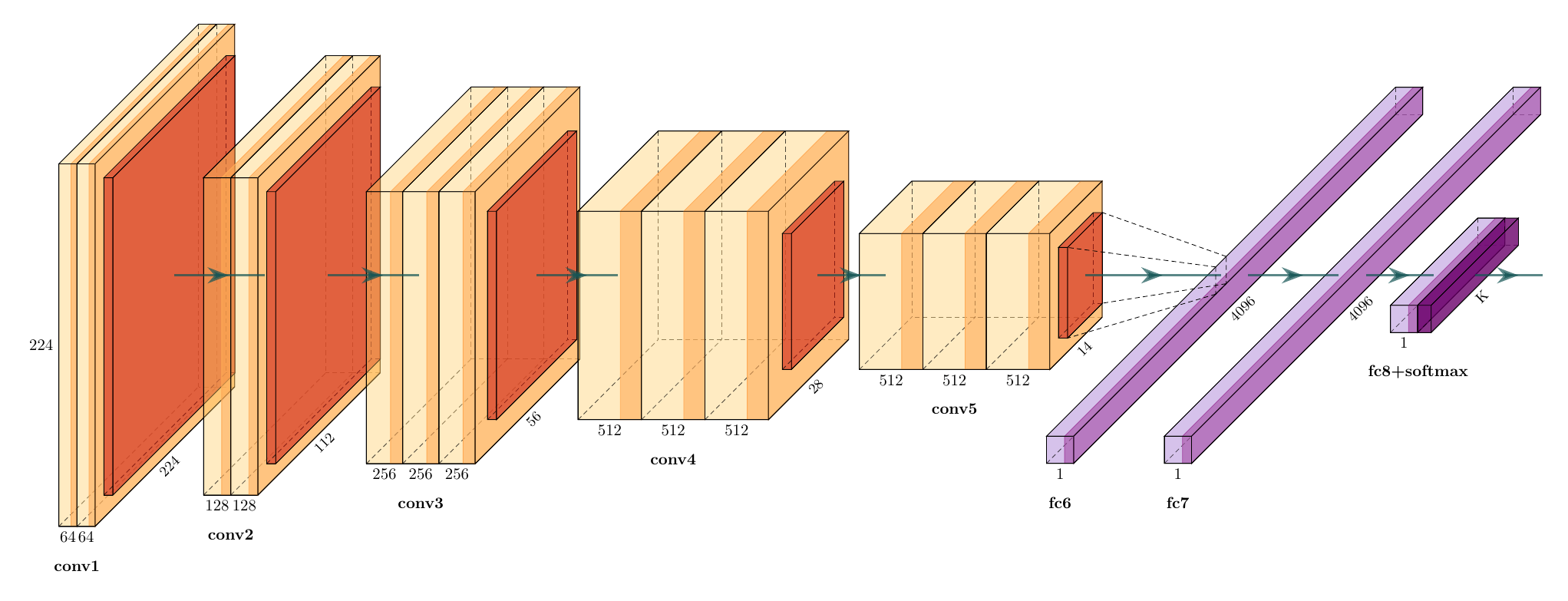}
    \caption{Schematic of a VGG16 network for ImageNet. The input data is convolved and down-sampled repeatedly. The final image classification is performed by flattening the image and feeding it to a classic feed-forward densely connected neural network. The 1000 output nodes for the 1000 ImageNet classes are normalized by a final softmax layer (cf.~equation~\ref{eq:softmax}). Visualization library \citep{haris_iqbal_2018_2526396}}
    \label{fig:vgg16}
\end{figure}

The first model to discuss is the VGG-16 model, a 16-layer deep convolutional neural network \citep{simonyan2014very} represented in figure~\ref{fig:vgg16}. This network was an attempt at building even deeper networks and uses small $3\times3$ convolutional filters in the network, called $f$ in equation~\ref{eq:convolution}. This small filter-size was sufficient to build powerful models that abstract the information from layer to deeper layer, which is easy to visualize and generalize well. The trained model on natural images also transfers well to other domains like seismic interpretation \citep{dramsch2018deep}. Later, the concept of Network-in-Network was introduced, which suggested defined sub-networks or blocks in the larger network structure \citep{lin2013network}. The ResNet architecture uses this concept of blocks to define residual blocks. These use a shortcut around a convolutional block \citep{he2016deep} to achieve neural networks with up to 152 layers that still generalize well. ResNets and residual blocks, in particular, are very popular in modern architectures including the shortcuts or skip connections they popularized, to address the following problem:

\begin{quote}
  When deeper networks start converging, a degradation problem has been exposed: with the network depth increasing, accuracy gets saturated (which might be unsurprising) and then degrades rapidly. Unexpectedly, such degradation is not caused by overfitting, and adding more layers to a suitably deep model leads to higher training error.
  \begin{flushright}\citet{he2016deep}\end{flushright}
\end{quote}

\begin{figure}
    \includegraphics[width=\textwidth]{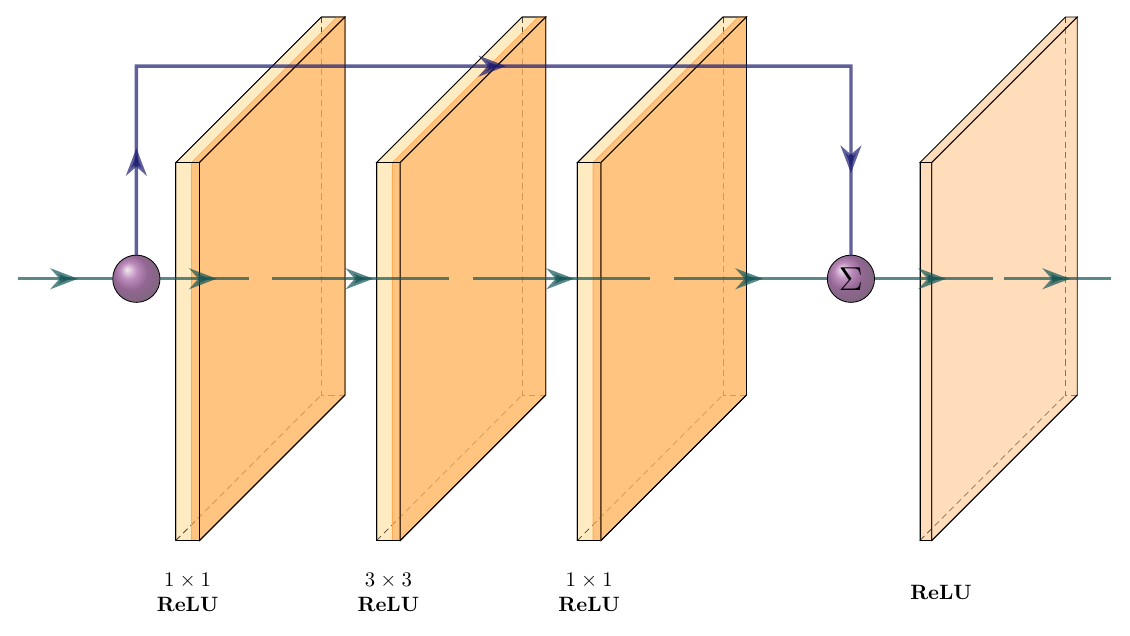}
    \caption{Schematic of a ResNet block. The block contains a $1\times1$, $3\times3$, and $1\times1$ convolution with ReLU activation. The output is concatenated with the input and passed through another ReLU activation function.}
    \label{fig:resnet}
\end{figure}

The developments and successes in image classification on benchmark competitions like ImageNet and Pascal-VOC inspired applications in automatic seismic interpretation. These networks are usually single image classifiers using convolutional neural networks (CNNs). The first application of a convolutional neural network to seismic data used a relatively small deep CNN for salt identification \citep{Waldeland2017-tx}. The open source software "MaLenoV" implemented a single image classification network, which was the earliest freely available implementation of deep learning for seismic interpretation \citep{malenov}. \citet{dramsch2018deep} applied pre-trained VGG-16 and ResNet50 single image seismic interpretation. Recent succesful applications build upon pre-trained pre-built architectures to implement into more sophisticated deep learning systems, e.g. semantic segmentation. Semantic segmentation is important in seismic interpretation. This is already a narrow field of application of machine learning and it can be observed that many early applications focus on sub-sections of seismic interpretation utilizing these pre-built architectures such as salt detection \citep{Waldeland2018-hj,Di2018-dz,Gramstad2018-ql}, fault interpretation \citep{araya2017automated, Guitton2018-gd, Purves2018-dy}, facies classification \citep{Chevitarese2018-kd, dramsch2018deep}, and horizon picking \citep{Wu2018-hg}. In comparison, this is however, already a broader application than prior machine learning approaches for seismic interpretation that utilized very specific seismic attributes as input to self-organizing maps (SOM) for e.g. sweet spot identification \citep{Guo2017-ij,Zhao2017-gv,Roden2017-pv}.

\begin{figure}[H]
    \centering
    
    \includegraphics[width=\textwidth]{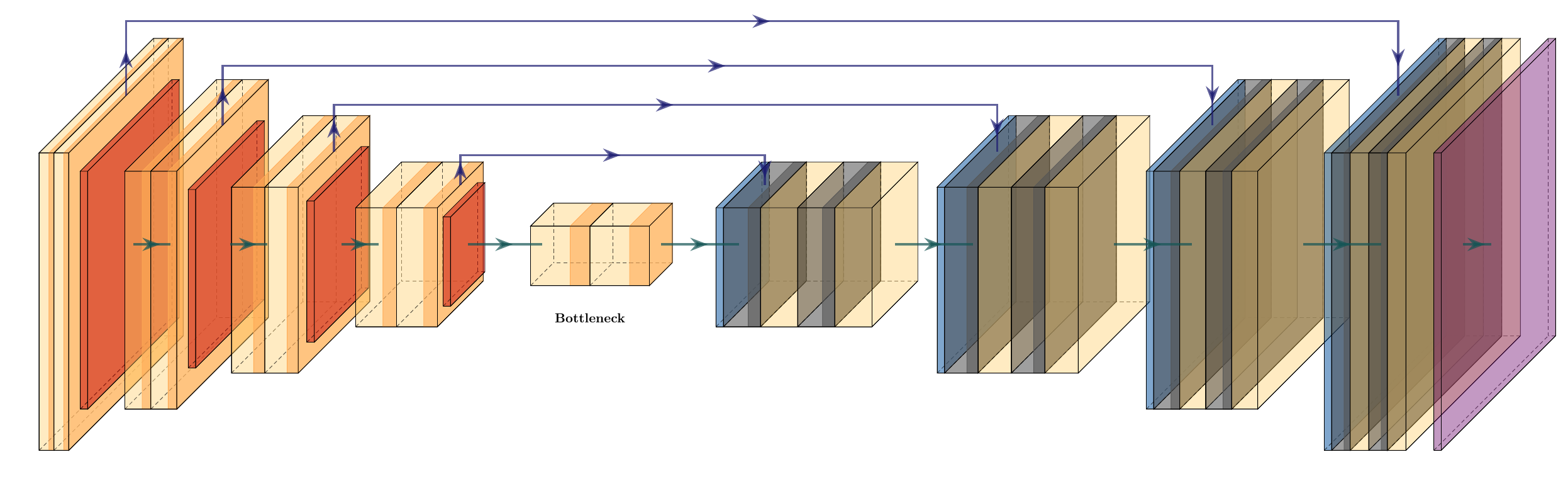}
    \caption{Schematic of Unet architecture. Convolutional layers are followed by a downsampling operation in the encoder. The central bottleneck contains a compressed representation of the input data. The decoder contains upsampling operations followed by convolutions. The last layer is commonly a softmax layer to provide classes. Equally sized layers are connected via shortcut connections.}
    \label{fig:unet}
\end{figure}

In geoscience single image classification, as presented in the ImageNet challenge, is less relevant than other applications like image segmentation and time series classification. The developments and insights resulting from the ImageNet challenge were, however, transferred to network architectures that have relevance in machine learning for science. Fully convolutional networks are a way to better achieve image segmentation. A particularly useful implementation, the U-net, was first introduced in biomedical image segmentation, a discipline notorious for small datasets \citep{ronneberger2015unet}. The U-net architecture shown in Figure~\ref{fig:unet} utilizes several shortcuts in an encoder-decoder architecture to achieve stable segmentation results. Shortcuts (or skip connections) are a way in neural networks to combine the original information and the processed information, usually through concatenation or addition. In ResNet blocks this concept is extended to an extreme, where every block in the architecture contains a shortcut between the input and output, as seen in Figure~\ref{fig:resnet}. These blocks are universally used in many architectures to implement deeper networks, i.e. ResNet-152 with 60 million parameters, with fewer parameters than previous architectures like VGG-16 with 138 million parameters. Essentially, enabling models that are ten times as deep with less than half the parameters, and significantly better accuracy on image benchmark problems.

In 2018 the seismic contractor TGS made a seismic interpretation challenge available on the data science competition platform Kaggle. Successful participants in the competition combined ResNet architectures with the Unet architecture as their base architecture and modified these with state-of-the-art image segmentation applications \citep{tgsSaltBodiesSegmentation2019}.  Moreover, \citet{dramsch2018deep} showed that transferring networks trained on large bodies of natural images to seismic data yields good results on small datasets, which was further confirmed in this competition. The learnings from the TGS Salt Identification challenge have been incorporated in production scale models that perform human-like salt interpretation \citep{sen2020saltnet}. In broader geoscience, U-nets have been used to model global water storage using GRAVE satellite data \citep{sun2019combining}, landslide prediction \citep{Hajimoradlou2019PredictingLU}, and earthquake arrival time picking \citep{Zhu2018-ma}. A more classical approach identifies subsea scale worms in hydrothermal vents \citep{shashidhara2020instance}, whereas \citet{dramsch20193dwarping} includes a U-net in a larger system for unsupervised 3D timeshift extraction from 4D seismic.

This modularity of neural networks can be seen all throughout the research and application of deep learning. New insights can be incorporated into existing architectures to enhance their predictive power. This can be in the form of swapping out the activation function $\sigma$ or including new layers for improvements e.g. regularization with batch normalization \citep{ioffe2015batch}. The U-net architecture originally is relatively shallow, but was modified to contain a modified ResNet for the Kaggle salt identification challenge instead \citep{tgsSaltBodiesSegmentation2019}. Overall, serving as examples for the flexibility of neural networks. 

\subsection{Generative Adversarial Networks}
Generative adversarial networks (GAN) take composition of neural network to another level, where two networks are trained in aggregate to get a desired result. In GANs, a generator network $G$ and a discriminator network $D$ work against each other in the training loop \citep{goodfellow2014generative}. The generator $G$ is set up to generate samples from an input, these were often natural images in early GANs, but has now progressed to anything from time series \citep{engel2019gansynth} to high-energy physics simulation \citep{paganini2018calogan}. The discriminator network $D$ attempts to distinguish whether the sample is generated from $G$ i.e. fake or a real image from the training data. Mathematically, this defines a min max game for the value function $V$ of $G$ and $D$
\begin{equation}
\min_G \max_D V (D, G) = \mathbb{E}_{x\sim p_{data}(x)} [\log D(x)] + \mathbb{E}_{z\sim p_z(z)} [\log(1 - D(G(z)))],
\end{equation}
with $x$ representing the data, $z$ is the latent space $G$ draws samples from, and $p$ represents the respective probability distributions. Eventually reaching a Nash equlibrium \citep{nash1951non}, where neither the generator network $G$ can produce better outputs, nor the discriminator network $D$ can improve its capability to discern between fake and real samples.

Despite how versatile U-nets are, they still need an appropriate defined loss function and labels to build a discriminative model. GANs however, build a generative model that approximates the training sample distribution in the Generator and a discriminative model of the Discriminator modeled dynamically through adversarial training. The Discriminator effectively providing an adversarial loss in a GAN. In addition to providing two models that serve different purposes, learning the training sample distribution with an adversarial loss makes GANs one of the most versatile models currently discovered. \citet{Mosser2017-ml} were applied GANs early on to geoscience, modeling 3D porous media at the pore scale with a deep convolutional GAN. The authors extended this approach to conditional simulations of oolithic digital rock \citep{Mosser2018-cr}. Early applications of GANs also included approximating the problem of velocity inversion of seismic data \citep{mosser2018rapid}, geostatistical inversion \citep{Laloy2017-lp}, and generating seismograms \citep{krischer2017generating}. \citet{Richardson2018-py} integrate the Generator of the GAN into full waveform inversion of the scalar wavefield. Alternatively, a Bayesian inversion using the Generator as prior for velocity inversion was introduced in \citet{Mosser2018-hm}. In geomodeling, generation of geological channel models was presented \citep{chan2017parametrization}, which was subsequently extended with the capability to be conditioned on physical measurements \citep{dupont2018generating}. Naturally, GANs were applied to the growing field of automatic seismic interpretation \citep{lu2018using}. 

\subsection{Recurrent Neural Network Architectures}
\begin{figure}
    \centering
    
    \includegraphics[width=\textwidth]{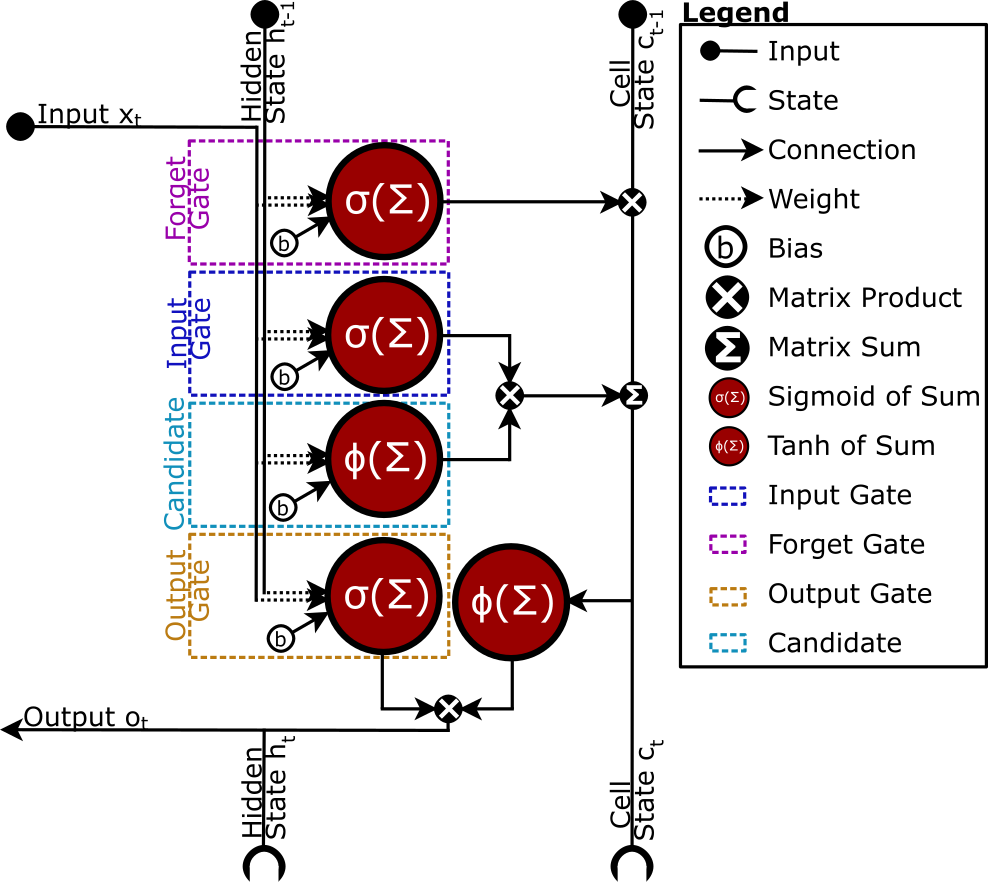}
    \caption{Schematic of LSTM architecture. The input data is processed together with the hidden state and cell state. The LSTM avoid the exploding gradient problem by implemented a input, forget, and output gate.}
    \label{fig:lstm}
\end{figure}

The final type of architecture applied in geoscience is recurrent neural networks (RNN). In contrast to all previous architectures, recurrent neural networks feed back into themselves. There are many types of RNNs, Hopfield networks being one that were  applied to seismic source wavelet prediction \citep{Wang1992-mq} early on. However, LSTMs \citep{hochreiter1997long} are the main application in geoscience and wider machine learning. This type of network achieves state-of-the-art performance on sequential data like language tasks and time series applications. LSTMs solve some common problems of RNNs by implementing specific gates that regulate information flow in an LSTM cell, namely, input gate, forget gate, and output gate, visualized in Figure~\ref{fig:lstm}. The input gate feeds input values to the internal cell. The forget gate overwrites the previous state. Finally, the output gate regulates the direct contribution of the input value to the output value combined with the internal state of the cell. Additionally, a peephole functionality helps with the training that serves as a shortcut between inputs and gates.

A classic application of LSTMs is text analysis and natural language understanding, which has been applied to geological relation extraction from unstructured text documents \citep{luo2017attention,Blondelle2017-dv}. Due to the nature of LSTMs being suited for time series data, it is has been applied to seismological event classification of volcanic activity \cite{titos2018detection}, multi-factor landslide displacement prediction \citep{xie2019application}, and hydrological modelling \citep{kratzert2019benchmarking}. \citet{talarico2019comparison} applied LSTM to model sedimentological sequences and compared the model to baseline Hidden Markov Model (HMM), concluding that RNNs outperform HMMs based on first-order Markov chains, while higher order Markov chains were too complex to calibrate satisfactorily. Gated Recurrent Unit (GRU) \citep{cho2014learning} is another RNN developed based on the insights into LSTM, which was applied to predict petrophysical properties from seismic data \citep{alfarraj2019petrophysical}.

The scope of this review only allowed for a broad overview of types of networks, that were successfully applied to geoscience. Many more specific architectures exist and are in development that provide different advantages. Siamese networks for one-shot image analysis \citep{koch2015siamese}, transformer networks that largely replaced LSTM and GRU in language modelling \citep{vaswani2017attention}, or attention as a general mechanism in deep neural networks \citep{zheng2017learning}. 

Neural network architectures have been modified and applied to diverse problems in geoscience. Every architecture type is particularly suited to certain data types that are present in each field of geoscience. However, fields with data present in machine-readable format experienced accelerated adoption of machine learning tools and applications. For example, \citet{Ross2018-kt} were able to successfully apply CNNs to seismological phase detection, relying on an extensive catalogue of hand-picked data \citep{Ross2018-rx} and consequently generalize this work \citep{ross2018generalized}. It has to be noted that synthetic or specifically sampled data can introduce an implicit bias into the network \citep{wirgin2004inverse,kim2019learning}. Nevertheless, particularly this blackbox property of machine learning model makes them versatile and powerful tools that were leveraged in every subdiscipline of the Earth sciences.

\subsection{The State of ML on Geoscience}
Overall, geoscience and especially geophysics has followed developments in machine learning closely. Across disciplines, machine learning methods have been applied to various problems that can generally be categorized into three sections:
\begin{enumerate} 
\item Build a surrogate ML model of a well-understood process. This model usually provides an advantage in computational cost. 
\item Build an ML model of a task previously only possible with human interaction, interpretation, or knowledge and experience. 
\item Build a novel ML model that performs a task that was previously not possible. 
\end{enumerate}

Granulometry on SEM images is an example of an application in category I, where previously sediments were hand-measured in images \citep{dramsch2018gaussian}. Applying large deformation diffeomorphic mapping of seismic data was computationally infeasible for matching 4D seismic data, however, made feasible by applying a U-net architecture to the problem of category II \citep{dramsch20193dwarping}. The problem of earthquake magnitude prediction falls into category III due to the complexity of the system but was nevertheless approached with neural networks \citep{panakkat2007neural}.

\begin{figure}
    \centering
    \includegraphics[width=\textwidth]{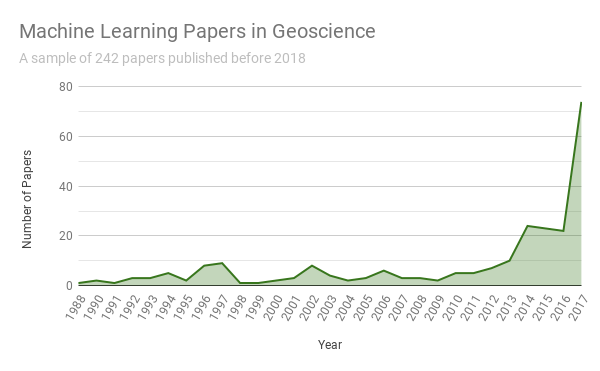}
    \caption{Bibliometry of 242 papers in Machine Learning for Geoscience per year. Search terms include variations of machine learning terms and geoscientific subdisciplines but exclude remote sensing and kriging.}
    \label{fig:number-papers}
\end{figure}

The accessibility of tools, knowledge, and compute make this cycle of machine learning enthusiasm unique, with regard to previous decades. This unprecedented access to tools makes the application of machine learning algorithms to any problem possible, where data is available. The bibliometrics of machine learning in geoscience, shown in figure~\ref{fig:number-papers} serve as a proxy for increased access. These papers include varying degrees of depth in application and model validation. One of the primary influences for the current increase in publications are new fields such as automatic seismic interpretation, as well as, publications soliciting and encouraging machine learning publications. Computer vision models were relatively straight forward to transfer to seismic interpretation tasks, with papers in this sub-sub-field ranging from single 2D line salt identification models with limited validation to 3D multi-facies interpretation with validation on a separate geographic area.

Geoscientific publishing can be challenging to navigate with respect to machine learning. While papers investigating the theoretical fundamentals of machine learning in geoscience exist, it is clear that the overwhelming majority of papers present applications of ML to geoscientific problems. It is complex to evaluate whether a paper is a case study or a methodological paper with an exemplary application to a specific data set. Despite the difficulty of most thorough applications of ML, "idea papers" exist that simply present an established algorithm to a problem in geoscience without a specific implementation or addressing the possible caveats. On the flip-side, some papers apply machine learning algorithms as pure regression models without the aim to generalize the model to other data. Unfortunately, this makes meta-analysis articles difficult to impossible. This kind of meta-analysis article, is commonly done in medicine and considered a gold-standard study, and would greatly benefit the geoscientific community to determine the efficacy of algorithms on sets of similar problems.

Analogous to the medical field, obtaining accurate ground truth data, is often impossible and usually expensive. Geological ground truth data for seismic data is usually obtained through expert interpreters. Quantifying the uncertainty of these interpretations is an active field of research, which suggest a broader set of experiences and a diverse set of sources of information for interpretation facilitate correct geological interpretation between interpreters \citep{bond2007you}. Radiologists tasked to interpret x-ray images showed similar decreases in both inter- and intra-interpreter error rate with more diverse data sources \cite{jewett1992potential}. These uncertainties in the training labels are commonly known as "label noise" and can be a detriment to building accurate and generalizable machine learning models. A significant portion of data in geoscience, however, is not machine learning ready. Actual ground truth data from drilling reports is often locked away in running text reports, sometimes in scanned PDFs. Data is often siloed and most likely proprietary. Sometimes the amount of samples to process is so large that many insights are yet to be made from samples in core stores or the storage rooms of museums. Benchmark models are either non-existent or made by consortia that only provide access to their members. Academic data is usually only available within academic groups for competitive advantage, respect for the amount of work, and fear of being exposed to legal and public repercussions. These problems are currently addressed by a culture change. Nevertheless, liberating data will be a significant investment, regardless of who will work on it and a slow culture change can be observed already.

Generally, machine learning has seen the fastest successes in domains where decisions are cheap (e.g. click advertising), data is readily available (e.g. online shops), and the environment is simple (e.g. games) or unconstrained (e.g. image generation). Geoscience generally is at the opposite of this spectrum. Decisions are expensive, be it drilling new wells or assessing geohazards. Data is expensive, sparse, and noisy. The environment is heterogeneous and constrained by physical limitations. Therefore, solving problems like automatic seismic interpretation see a surge of activity having fewer constraints initially. Problems like inversion have solutions that are verifiably wrong due to physics. These constraints do not prohibit machine learning applications in geoscience. However, most successes are seen in close collaboration with subject matter experts. Moreover, model explainability becomes essential in the geoscience domain. While not being a strict equivalency, simpler models are usually easier to interpret, especially regarding failure modes. 

A prominent example of "excessive" \citep{mignan2019deeper} model complexity was presented in \citet{devries2018deep} applying deep learning to aftershock prediction. Independent data scientists identified methodological errors, including data leakage from the train set to the test set used to present results \citep{aftershockissues}. Moreover, \citet{mignan2019one} showed that using the central physical interpretation of the deep learning model, using the von Mises yield criterion, could be used to build a surrogate logistic regression. The resulting surrogate or baseline model outperforms the deep network and overfits less. Moreover, replacing the \textasciitilde13,000 parameter model with the two-parameter baseline model increases calculation speed, which is essential in aftershock forecasting and disaster response\footnote{All authors point out the potential in deep and machine learning research in geoscience regardless and do not wish to stifle such research.\citep{aftershockissues,mignan2019one}}. More generally, this is an example where data science practices such as model validation, baseline models, and preventing data leakage and overfitting become increasingly important when the tools of applying machine learning become readily available.

Despite potential setbacks and the field of deep learning and data science being relatively young, they can rely on mathematical and statistical foundations and make significant contributions to science and society. Machine learning systems have contributed to modelling the protein structure of the current pandemic virus COVID-19 \citep{alphacovid}. A deep learning computer vision system was built to stabilize food safety by identifying Cassava plant disease on offline mobile devices \citep{ramcharan2017deep,ramcharan2019mobile}. Self-driving cars have become a possibility \citep{bojarski2016end} and natural language understanding has progressed significantly \citep{devlin2018bert}.

Geoscience is slower in the adoption of machine learning, compared to other disciplines. To be able to adapt the progress in machine learning research, many valuable data sources have to be made machine-readable. There has already been a change in making computer code open source, which has lead to collaborations and accelerating scientific progress. While specific open benchmark data sets have been tantamount to the progress in machine learning, it is questionable whether these would be beneficial to machine learning in geoscience. The problems are often very complex with non-unique explanations and solutions, which historically has lead to disagreements over geophysical benchmark data sets. Open data and open-source software, however, have and will play a significant role in advancing the field. Examples of this include basic utility function to load geoscientific data \citep{segyio} or more specifically cross-validation functions tailored to geoscience \citep{uieda2018}. 

Moreover, machine learning is fundamentally conservative, training on available data. This bias of data collection will influence the ability to generate new insights in all areas of geoscience. Machine learning in geoscience may be able to generate insights and establish relationships in existing data. Entirely new insights from previously unseen or analysis of particularly complex models will still be a task performed by trained geoscientists. Transfer learning is an active field of machine learning research, that geoscience can significantly benefit from. However, no significant headway has been made to transfer trained machine learning models to out-of-distribution data, i.e. data that is conceptually similar but explicitly different from the training data set. The fields of self-supervised learning, including reinforcement learning that can learn by exploration, may be able to approach some of these problems. They are, however, notoriously hard to set up and train, necessitating significant expertise in machine learning.

Large portions of publications are concerned with weakly or unconstrained predictions such as seismic interpretation and other applications that perform image recognition on SEM or core photography. These methods will continue to improve by implementing algorithmic improvements from machine learning research, specialized data augmentation strategies, and more diverse training data being available. New techniques such as multi-task learning \citep{kendall2018multi} which improved computer vision and computer linguistic models, deep bayesian networks \citep{mosser2019probabilistic} to obtain uncertainties, noisy teacher-student networks \citep{xie2019self} to improve training, and transformer networks \citep{graves2012sequence} for time series processing, will significantly improve applications in geoscience. For example, automated seismic interpretation may advance to provide reliable outputs for relatively difficult geological regimes beyond existing solutions. Success will be reliant on interdisciplinary teams that can discern why geologically specific faults are important to interpret, while others would be ignored in manual interpretations, to encode geological understanding in automatic interpretation systems. 

Currently, the most successful applications of machine learning and deep learning, tie into existing workflows to automate sub-tasks in a grander system. These models are highly specific, and their predictive capability does not resemble an artificial intelligence or attempt to do so. Mathematical constraints and existing theory in other applied fields, especially neuroscience, were able to generate insights into deep learning and geoscience has the opportunity to develop significant contributions to the area of machine learning, considering their unique problem set of heterogeneity, varying scales and non-unique solutions. This has already taken place with the wider adoption of "kriging" or more generally Gaussian processes into machine learning. Moreover, known applications of signal theory and information theory employed in geophysics are equally applicable in machine learning, with examples utilizing complex-valued neural networks \citep{trabelsi2017deep}, deep Kalman filters \citep{krishnan2015deep}, and Fourier analysis \citep{tancik2020fourier}. Therefore, possibly enabling additional insights, particularly when integrated with deep learning, due to its modularity and versatility.

Previous reservations about neural networks included the difficulty of implementation and susceptibility to noise in addition to computational costs. Research into updating trained models and saving the optimizer state with the model has in part alleviated the cost of re-training existing models. Moreover, fine-tuning pre-trained large complex models to specific problems has proven successful in several domains. Regularization techniques and noise modelling, as well as data cleaning pipelines, can be implemented to lessen the impact of noise on machine learning models. Specific types of noise can be attenuated or even used as an additional source of information. The aforementioned concerns have mainly transitioned into a critique about overly complex models that overfit the training data and are not interpretable. Modern software makes very sophisticated machine learning models, and data pipelines available to researchers, which has, in turn, increased the importance to control for data leakage and perform thorough model validation.

Currently, machine learning for science primarily relies on the emerging field of explainability \citep{lundberg2018explainable}. These provide primarily post-hoc explanations for predictions from models. This field is particularly important to evaluate which inputs from the data have the strongest influence on the prediction result. The major point of critique regarding post-hoc explanations is that these methods attempt to explain how the algorithm reached a wrong prediction with equal confidence. Bayesian neural networks intend to address this issue by providing confidence intervals for the prediction based on prior beliefs. These neural networks intend to incorporate prior expert knowledge into neural networks, which can be beneficial in geoscientific applications, where strong priors can be necessary. Machine learning interpretability attempts to impose constraints on the machine learning models to make the model itself explainable. Closely related to these topics is the statistics field of causal inference. Causal inference attempts to model the cause of variable, instead of correlative prediction. Some methods exist that can perform causal machine learning, i.e. causal trees \citep{athey2016recursive}. These three fields will be necessary to glean verifiable scientific insights from machine learning in geoscience. They are active fields of research and more involved to correctly apply, which often makes cooperation with a statistician necessary.

In conclusion, machine learning has had a long history in geoscience. Kriging has progressed into more general machine learning methods, and geoscience has made significant progress applying deep learning. Applying deep convolutional networks to automatic seismic interpretation has progressed these methods beyond what was possible, albeit still being an active field of research. Using modern tools, composing custom neural networks, and conventional machine learning pipelines has become increasingly trivial, enabling wide-spread applications in every sub-field of geoscience. Nevertheless, it is important to acknowledge the limitations of machine learning in geoscience. Machine learning methods are often cutting-edge technology, yet properly validated models take time to develop, which is often perceived as inconvenient when working in a hot scientific field. Despite being cutting edge, it is important to acknowledge that none of these applications are fully automated, as would be suggested by the lure of artificial intelligence. Nevertheless, within applied geoscience, significant new insights have been presented. Applications in geoscience are using machine learning as a utility for data pre-processing, implementing previous insights beyond the theory and synthetic cases, or the model itself enabling unprecedented applications in geoscience. Overall, applied machine learning has matured into an established tool in computational geoscience and has the potential to provide further insights into the theory of geoscience itself.

\section*{References}

\bibliography{bib}





\end{document}